\documentclass[twocolumn,trackchanges,twocolappendix]{aastex701}
\newcommand{\exotr}{\mbox{\texttt{ExoTR}}}
\usepackage{multirow}
\usepackage{threeparttable}
\usepackage[version=4]{mhchem}
\usepackage{enumitem}
\usepackage{soul}
\usepackage{xcolor}
\usepackage{subcaption}
\usepackage{array}

\newcommand\planetname{TOI-199\,b}

\newif\ifarxiv
\arxivfalse        

\ifarxiv
  \usepackage{fancyhdr}
  \fancypagestyle{firstpage}{%
    \fancyhf{}%
    \fancyfoot[C]{\footnotesize \textcopyright\ 2025. All rights reserved.}%
  }

  \makeatletter
  \let\origmaketitle\maketitle
  \def\maketitle{\origmaketitle\thispagestyle{firstpage}}
  \makeatother
\fi

\begin{document}

\title{Methane on the temperate exo-Saturn TOI-199\,b}

\correspondingauthor{Aaron Bello-Arufe}
\email{aaron.bello.arufe@jpl.nasa.gov}

\author[0000-0003-3355-1223]{Aaron Bello-Arufe}
\affiliation{Jet Propulsion Laboratory, California Institute of Technology, Pasadena, CA 91109, USA}
\email{aaron.bello.arufe@jpl.nasa.gov}

\author[0000-0003-2215-8485]{Renyu Hu}
\affiliation{Jet Propulsion Laboratory, California Institute of Technology, Pasadena, CA 91109, USA}
\affiliation{Department of Astronomy \& Astrophysics, The Pennsylvania State University, University Park,
PA 16802, USA}
\affiliation{Center for Exoplanets and Habitable Worlds, The Pennsylvania State University, University Park,
PA 16802, USA}
\affiliation{Institute for Computational and Data Science, The Pennsylvania State University, University Park,
PA 16802, USA}
\email{rqh5611@psu.edu}

\author[0000-0002-8749-823X]{Mantas Zilinskas}
\affiliation{Jet Propulsion Laboratory, California Institute of Technology, Pasadena, CA 91109, USA}
\email{mantas.zilinskas@jpl.nasa.gov}

\author[0000-0002-1551-2610]{Jeehyun Yang}
\affiliation{The University of Chicago, Chicago, IL 60637, USA}
\affiliation{Jet Propulsion Laboratory, California Institute of Technology, Pasadena, CA 91109, USA}
\email{jeehyuny@uchicago.edu}

\author[0000-0002-4675-9069]{Armen Tokadjian}
\affiliation{Jet Propulsion Laboratory, California Institute of Technology, Pasadena, CA 91109, USA}
\email{armen.tokadjian@jpl.nasa.gov}

\author[0000-0003-0156-4564]{Luis Welbanks}
\affiliation{School of Earth and Space Exploration, Arizona State University, 781 Terrace Mall, Tempe, AZ 85287, USA}
\email{luis.welbanks@asu.edu}

\author[0000-0002-3263-2251]{Guangwei Fu}
\affiliation{Department of Physics \& Astronomy, Johns Hopkins University, Baltimore, MD 21218 USA}
\email{guangweifu@gmail.com}

\author[0000-0002-0371-1647]{Michael Greklek-McKeon}
\affiliation{Earth and Planets Laboratory, Carnegie Institution for Science, Washington, DC 20015, USA}
\email{michael@caltech.edu}

\author[0000-0002-1830-8260]{Mario Damiano}
\affiliation{Jet Propulsion Laboratory, California Institute of Technology, Pasadena, CA 91109, USA}
\email{mario.damiano@jpl.nasa.gov}

\author[0000-0002-0672-9658]{Jonathan Gomez Barrientos}
\affiliation{Division of Geological and Planetary Sciences, California Institute of Technology, Pasadena, CA 91125, USA}
\email{jgomezba@caltech.edu}

\author[0000-0002-5375-4725]{Heather A. Knutson}
\affiliation{Division of Geological and Planetary Sciences, California Institute of Technology, Pasadena, CA 91125, USA}
\email{hknutso2@caltech.edu}

\author[0000-0001-6050-7645]{David K. Sing}
\affiliation{Department of Earth \& Planetary Sciences, Johns Hopkins University, Baltimore, MD 21218, USA}
\affiliation{Department of Physics \& Astronomy, Johns Hopkins University, Baltimore, MD 21218 USA}
\email{dsing@jhu.edu}

\author[0000-0002-8706-6963]{Xi Zhang}
\affiliation{Department of Earth and Planetary Sciences, University of California Santa Cruz, Santa Cruz, CA 95064, USA}
\email{xiz@ucsc.edu}

\begin{abstract}
Temperate ($T_{\rm eq}<400$~K) gas giants represent an unexplored frontier in exoplanet atmospheric spectroscopy. Orbiting a G-type star every $\sim100$ days, the Saturn-mass exoplanet TOI-199\,b ($T_{\rm eq}=350$~K) is one of the most favorable low-temperature gas giants for atmospheric study. Here, we present its transmission spectrum from a single transit observed with JWST’s NIRSpec G395M mode. Despite lower-than-nominal precision due to a pointing misalignment, the spectrum reveals the presence of CH$_4$ (Bayes factor of $\sim$700 in a cloudy atmosphere), corresponding to a metallicity of $\rm{C/H}=13^{+78}_{-12}\times$ solar, although the absence of detectable CO and CO$_2$ at the current precision disfavors metallicities $\gtrsim50\times$ solar. We also tested several haze prescriptions (Titan-like tholin, soot, and water-rich tholin), but the preference for these models is weak (Bayes factors of $\sim 2$ relative to the clear case). The spectrum also shows an increase in transit depth near 3~$\mu$m, which our self-consistent models attribute to either NH$_3$ or, less likely, HCN. Follow-up observations could distinguish between these species, helping determine the planet's vertical mixing regime. The TOI-199 system exhibits strong transit timing variations (TTVs) due to an outer non-transiting giant planet. For planet c, our TTV analysis reduces its mass uncertainty by 50\% and prefers a slightly longer orbital period (but still within the conservative habitable zone) and higher eccentricity relative to previous studies. TOI-199\,b serves as the first data point for studying clouds and hazes in temperate gas giants, with the detection of spectral features in its transmission spectrum indicating that temperate gas giants are promising targets for detailed atmospheric characterization.

\end{abstract}




\section{Introduction} 
Hot Jupiters have dominated the landscape of atmospheric studies of exoplanets over the past two decades. Their frequent transits, large atmospheric scale heights, and favorable orbital geometry make them the most accessible exoplanets to characterize. These planets have been extensively observed with HST, Spitzer, and ground-based telescopes \citep[e.g.,][]{knutson2007,sing2016,ehrenreich2020}, and more recently with JWST \citep[e.g.,][]{alderson2023early,feinstein2023early}, with recent milestones including the detection of quartz clouds and photochemically produced SO$_2$ in their atmospheres \citep{tsai2023photochemically,grant2023jwst}.

By contrast, low-temperature ($T_{\rm eq}<400$ K) transiting giants remain unexplored. A handful of these planets are known to transit their host stars, and some are highly suitable for atmospheric characterization via transmission spectroscopy. Many of these planets orbit G-type stars on long-period ($\gtrsim100$ days) orbits \citep[e.g.,][]{mancini2016,beichman2018} and, unlike the vast majority of exoplanets probed with transit spectroscopy, they are not expected to be tidally locked \citep{showman2015}. As a result, longitudinal temperature gradients in their atmospheres should be orders of magnitude smaller than for hot Jupiters.

At temperatures intermediate between those of Jupiter/Saturn and those of hot Jupiters, we expect to see unique manifestations of atmospheric chemistry. At $T_{\rm eq}<400$ K, CH$_4$ is the dominant carbon-bearing molecule, and a substantial fraction of nitrogen is expected to be in NH$_3$ \citep{fortney2020beyond,hu2021photochemistry,ohno2023nitrogen}, with many low-temperature giant exoplanets also remaining warm enough to avoid H$_2$O condensation. This trait makes temperate gas giants ideal targets to measure the atmospheric abundances of C, N, and O and constrain their feeding zones and migration histories \citep[e.g.,][]{oberg2011,mordasini2016imprint,espinoza2017metal,ali2017disentangling,cridland2019connecting,cridland2020connecting}. Moreover, photochemical processes that drive hydrocarbon formation in the Jovian atmosphere \citep{gladstone1996hydrocarbon,moses2005} instead result in the formation of HCN over hydrocarbons in warmer gas giants \citep{hu2021photochemistry}, with quantitative predictions depending on $K_{\rm zz}$ and chemical networks \citep{yu2021identify,tsai2021inferring}. The mechanisms to form HCN have been studied extensively for Titan \citep[e.g.,][]{vuitton2019simulating}, early Earth \citep[e.g.,][]{airapetian2016prebiotic,rimmer2019hydrogen}, and hot Jupiters \citep[e.g.,][]{line2011thermochemical,kawashima2018theoretical,hobbs2019chemical}. Low-temperature giant exoplanets thus fill the temperature gap between Titan and hot Jupiters and can provide valuable empirical constraints on the formation of HCN on the early Earth.

With a Transmission Spectroscopy Metric of 107, the newly confirmed planet \planetname\ ($0.810\pm0.005\,{R}_{{\rm{J}}}$, \citealt{hobson2023}) stands out as one of the most favorable low-temperature giant planets for atmospheric characterization. Its mass and radius suggest a Saturn-like internal structure with an H$_2$-dominated atmosphere. The planet orbits its G9V host star every 105 days, receiving $2.5$ times Earth's irradiation, which corresponds to a zero-albedo equilibrium temperature of 352 K. Its relatively low mass ($0.17\pm0.02\,{M}_{{\rm{J}}}$) and eccentricity ($0.09_{-0.02}^{+0.01}$) suggest a low internal heat flux \citep{fortney2020beyond} and further solidify TOI-199\,b as a canonical test case for temperate giant exoplanets.

Here we present the first atmospheric reconnaissance of \planetname\ from a single transit observed with JWST/NIRSpec G395M. Establishing \planetname\ as the representative temperate gas giant for detailed atmospheric studies, this paper not only provides the initial constraints on its atmospheric properties, but also uses advanced atmospheric models to chart the next steps in the characterization. In addition, we analyze the JWST and TESS light curves to model the transit timing variations (TTVs) in the system, with the aim of refining the physical and orbital properties of the two known planets in the system, and predicting the transit times through to the end of the year 2040.

The paper is structured as follows. In Section~\ref{sec:observations_and_data_reduction}, we describe our observations and our two independent data-reduction pipelines, and we present the resulting transmission spectrum. Section~\ref{sec:retrievals} reports the results from our two independent retrieval analyses. In Section~\ref{sec:self-consistent}, we present a series of self-consistent photochemical models anchored by the current data to delineate the current constraints and additional information about the planet's atmosphere that could be obtained with improved spectral precisions. In Section~\ref{sec:ttvs} we present the TTV analysis. Finally, in Section~\ref{sec:discussion_and_conclusion}, we discuss the results and outline how future observations can be used to place constraints on the efficacy of vertical mixing and photochemistry in this archetype low-temperature giant exoplanet.

\section{Observations and Data Reduction}\label{sec:observations_and_data_reduction} 
As part of JWST's General Observer program 5177 (P.I. Hu), we observed a transit of \planetname\ on Dec 23, 2024 UTC. We collected the data using the Near InfraRed Spectrograph \citep[NIRSpec,][]{jakobsen2022,birkmann2022} with the medium-resolution grating G395M ($R\sim 1000$), the SUB2048 subarray, and seven groups per integration.

According to the JWST Exposure Time Calculator (ETC), the TOI-199 system ($m_J = 9.3$) is bright enough to saturate the NIRSpec wide aperture target acquisition (WATA). We instead chose a nearby infrared source as the acquisition target. However, due to target acquisition failure, the nearby source fell outside the aperture during the observations, and instead the WATA found a noise peak of $\sim80$ counts in the centroiding box (at least $\sim 150$ are typically required). As a result, our observations plausibly picked up the far wings of the point-spread function (PSF) of the science target or a diffraction spike. However, the differential nature of the transit measurement enabled us to still extract a useful transmission spectrum. The resulting spectral uncertainties are $\sim4-5\times$ larger than predicted by PandExo \citep{batalha2017}.

To ensure that our results are robust against different data treatment approaches, we performed two independent data reductions, which we detail below.

\subsection{\texttt{Eureka!} reduction}\label{subsubsec:eureka_reduction}
We reduced the data using version 1.1 of the \texttt{Eureka!} pipeline \citep{bell2022eureka}. \texttt{Eureka!} is structured into six stages: the first two stages calibrate the raw data; stage three performs optimal extraction; stages four and five generate and fit the lightcurves, respectively; and stage six displays the results. Our \texttt{Eureka!} setup was informed by prior applications to other NIRSpec datasets \citep{damiano2024,belloarufe2025,hu2025}, but with modifications to minimize the noise in the resulting lightcurves and including additional steps to mitigate the systematics introduced by the failed WATA, all of which are described below.

Starting from the uncalibrated raw data in the NRS1 detector (i.e., the \texttt{*nrs1\_uncal.fits} files), we ran stages one and two of \texttt{Eureka!}. These stages call a series of steps from the \texttt{jwst} pipeline (version 1.15.1, \citealt{bushouse2024}) to perform initial processing and calibration. We ran all the default steps for NIRSpec time-series observations (TSO) data, except for the flat-field and photometric calibration steps. For the jump step, which identifies ``up-the-ramp" outliers in each pixel, we increased the sigma threshold from the default value of 4 to a more conservative value of 7. To correct the superbias, we used the mean scale factor over all integrations. Prior to fitting the ramps, we performed group-level background subtraction using the average value in each detector column, excluding a region with a half-width of 8 pixels centered around the trace.

After calibration of the raw files, we ran \texttt{Eureka!}'s stage 3. Here, we extracted the detector columns 686--2044, which contain the spectral trace. We masked pixels with an odd data quality (DQ) entry, and we corrected the curvature of the trace by shifting each detector column by a whole number of pixels. To determine the position of the source, we fit a Gaussian function to each detector column. We applied another round of column-by-column background subtraction, this time using the average value of pixels located at least 9 pixels away from the center of the trace, with a 5$\sigma$ threshold for outlier rejection. For the optimal extraction of the spectra \citep{Horne1986}, we used an aperture with a half-width of 3 pixels and constructed the spatial profile using the median frame after clipping 5$\sigma$ outliers and smoothing it with a window length of 13. We then masked three excessively noisy detector columns: 1488, 1661, and 1798.

In stage four, we generated spectroscopic lightcurves from 2.87 to 5.17~$\mu \rm{m}$, with a bin size of $\Delta\lambda = 0.004~\rm{\mu m}$, which closely matches the native spectral resolution of NIRSpec G395M ($R\sim 1000$). However, in order to test the robustness of our results against different binning schemes, we produced a second set of lightcurves at $\Delta\lambda = 0.025~\rm{\mu m}$. We cleaned each lightcurve using a rolling boxcar filter with a width of 10 data points that recursively clipped 5$\sigma$ outliers, up to 20 iterations.

\begin{figure}
\centering
\includegraphics[width=\linewidth]{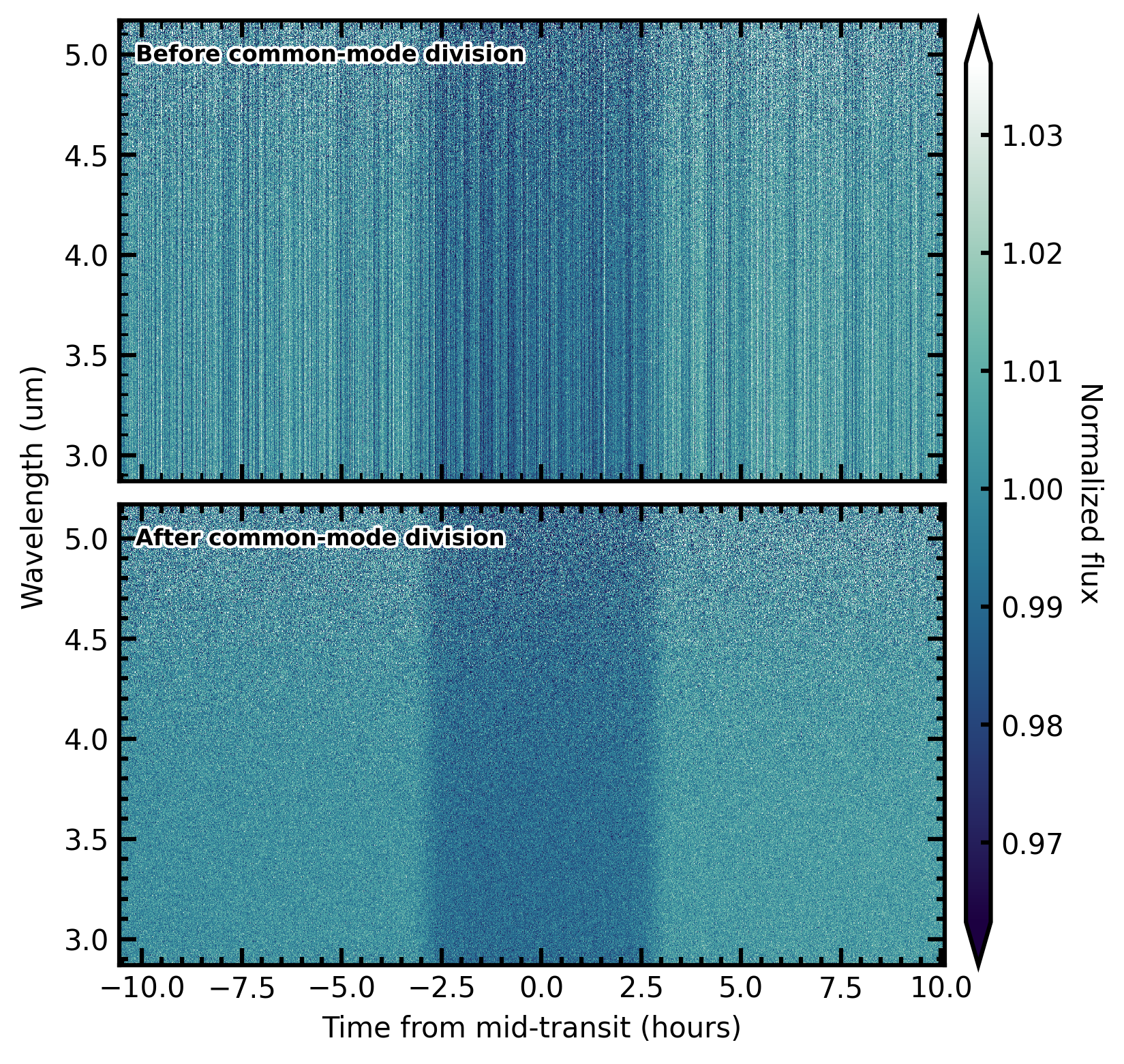}
\caption{Raw spectroscopic lightcurves, as extracted with \texttt{Eureka!} and binned to $\Delta\lambda = 0.004~\mu$m, before and after dividing out the common-mode noise model.}\label{fig:spectroscopic_lightcurves}
\end{figure}

The top panel in Figure~\ref{fig:spectroscopic_lightcurves} shows the raw spectroscopic lightcurves. While the transit of \planetname\ is clearly visible around the middle of the observations, the lightcurves show wavelength-correlated noise. This systematic noise is caused by the failed WATA and, fortunately, is achromatic. As shown in Figure~\ref{fig:spectroscopic_lightcurves}, we can effectively correct this noise by dividing out each spectroscopic lightcurve by a common-mode noise model. 

To construct the common-mode noise model, we divided the white lightcurve by the best-fit model, which consisted of a \texttt{batman} transit model \citep{kreidberg2015batman} multiplied by a systematics model \citep[see e.g.,][]{murgas2019,fu2024}. In our transit model, we fixed the period $P$, eccentricity $e$, and argument of periastron $\omega$ to the values reported by \citet{hobson2023}, and we assigned uniform priors to the remaining transit and orbital parameters, namely the planet-to-star radius ratio $R_p/R_\star$, the transit time $T_{0}$, the orbital inclination $i_p$, and the scaled semimajor axis $a/R_\star$. We also fitted for the quadratic limb-darkening coefficients ($q_1,q_2$, \citealt{kipping2013limbdark}), to which we assigned uniform priors between 0 and 1. Meanwhile, the systematics model consisted of a linear polynomial in time, and we also included a white noise multiplier to boost the error bars to match the scatter of the residuals. We performed the white lightcurve fits using the dynamic nested sampling package \texttt{dynesty} \citep{speagle2020}, which is included in \texttt{Eureka!}'s stage five. We used 2000 live points and set a convergence tolerance of $\Delta\log Z < 0.001$. The goal of this initial white lightcurve fit is solely to generate the common-mode noise model that is subsequently divided out from each spectroscopic lightcurve.

\begin{figure*}
\centering
\includegraphics[width=\linewidth]{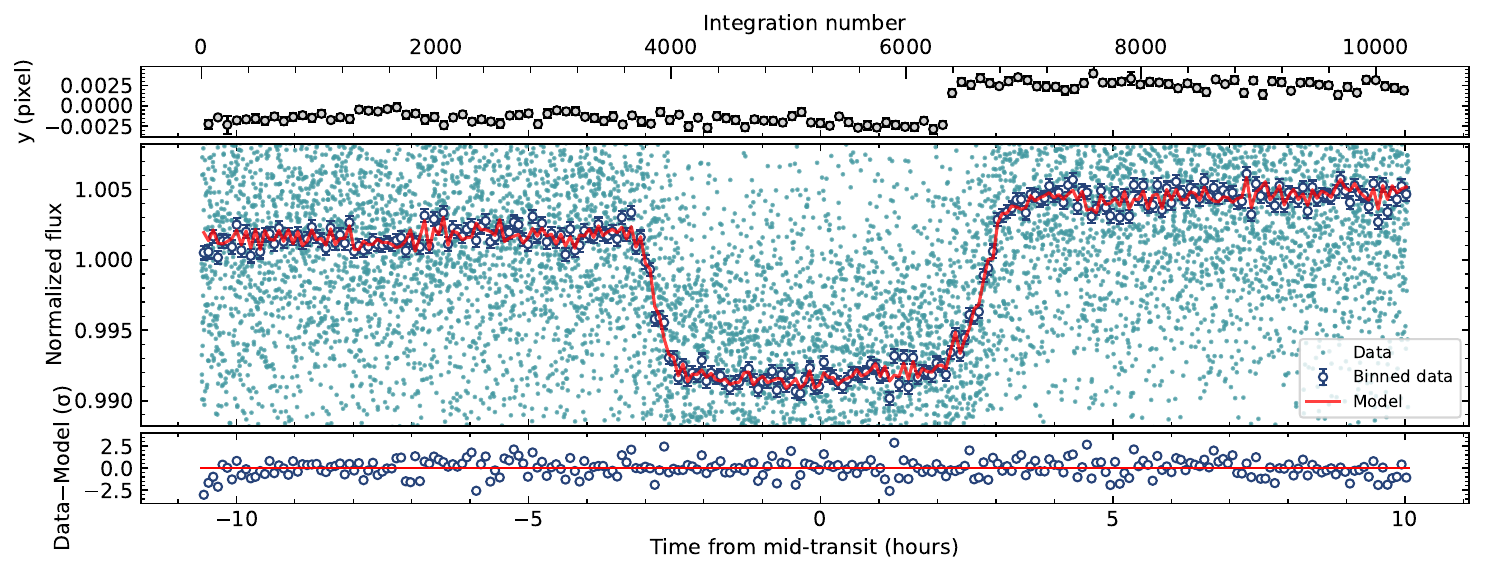}
\caption{\textit{Top:} mean-subtracted position of the trace along the $y$ (i.e., cross-dispersion) direction, binned by a factor of 80 to more easily visualize the jump occurring right before egress. \textit{Middle:} raw and binned ($\times 40$) white lightcurves of \planetname 's transit, observed with NIRSpec G395M, including the best-fit model, as extracted with \texttt{Eureka!}. \textit{Bottom:} residuals from the best-fit model, measured in $\sigma$.}\label{fig:white_lightcurve}
\end{figure*}

\begin{figure}
\centering
\includegraphics[width=\linewidth]{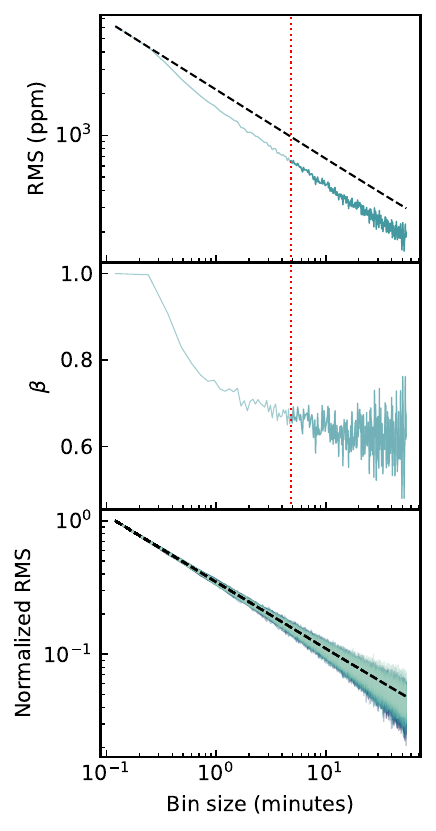}
\caption{\textit{Top:} RMS of the white lightcurve residuals as a function of bin size (solid teal line) and scaling expected for purely white noise (dashed black line). The dotted red line indicates the bin size used in the \texttt{Eureka!} white lightcurve fit (i.e., 40$\times$). \textit{Middle:} ratio of the two lines in the top plot (i.e., red noise factor, \citealt{winn2008}). \textit{Bottom:} normalized RMS of the residuals of the spectroscopic lightcurves from \texttt{Eureka!} as a function of bin size (solid colored lines) and scaling expected for purely white noise (dashed black line).}\label{fig:allan}
\end{figure}

We then produced a second white lightcurve fit, which we used to report the system parameters and fix them in the spectroscopic lightcurve fits. One of the differences between this white lightcurve fit and the previous one is that our new systematics model additionally includes linear decorrelation against the $x$ and $y$ position and the width of the trace. This choice is motivated by an improvement in the Bayesian evidence. Another difference is that we now also include a step function to account for the jump in the \textit{y} position (i.e., cross-dispersion direction) of the trace happening around the 6340th integration, right before egress (see Figure~\ref{fig:white_lightcurve}), which results in a jump in the lightcurve flux. Indeed, we found that linear decorrelation alone was not sufficient to account for the full extent of the jump in flux, and Bayesian model comparison favors including a step function in the fits. We assigned a uniform prior to the step time and a normal prior centered around 0 to the step size. The third and final difference is binning of the lightcurve. The rms of the residuals decreases more steeply than the $N^{-1/2}$ scaling expected for uncorrelated noise, indicating the presence of high-frequency correlations (Figure~\ref{fig:allan}, top panel). We binned the white lightcurve in time by a factor of 40 to mitigate these correlations, while still retaining sufficient points across the ingress and egress to accurately resolve the transit shape (Figure~\ref{fig:allan}, middle panel). The prior and posterior system parameter distributions from this fit are presented in Table \ref{table:bestfitwlc}, and the best-fit model is shown in Figure~\ref{fig:white_lightcurve}.

\begin{table}[]
\caption{Prior and Posterior System Parameter Distributions for \texttt{Eureka!}'s White Lightcurve Fit.}
\centering
\begin{tabular}{lcc}
\hline\hline
\textbf{Parameter}  & \textbf{Prior} & \textbf{Posterior} \\
\hline
$T_0$ (BJD$_{\rm{TDB}}–$2\,460\,668) & $\mathcal{U}$(0.15, 0.25) & 0.21244$^{+0.00039}_{-0.00039}$ \\
$i_p$ ($^\circ$) & $\mathcal{U}$(89.25, 90) & 89.737$^{+0.024}_{-0.021}$\\ 
$a/R_s$ & $\mathcal{U}$(50, 450) & 119.1$^{+3.0}_{-2.8}$\\
$P$ (days) & 104.854 (fixed) & 104.854 (fixed) \\
$e$ & 0.09 (fixed) & 0.09 (fixed) \\
$\omega$ & 350 (fixed) & 350 (fixed) \\
$R_p/R_s$ & $\mathcal{U}$(0.05, 0.15) & 0.10268$^{+0.00064}_{-0.00068}$\\
$q_1$  & $\mathcal{U}$(0, 1) & 0.028$^{+0.030}_{-0.016}$\\
$q_2$ & $\mathcal{U}$(0, 1) & 0.29$^{+0.39}_{-0.22}$\\ \hline
\end{tabular}\label{table:bestfitwlc}
\tablecomments{$\mathcal{U}(a,b)$ is the uniform distribution between values $a$ and $b$. The posterior values are reported as the median, with uncertainties given by the 16$^{\textup{th}}$ and 84$^{\textup{th}}$ percentiles.}
\end{table}

To generate the transmission spectrum, we fitted the spectroscopic lightcurves in the same manner as the second white lightcurve fit, but with the following modifications. First, we kept $T_0$, $i_p$, and $a/R_s$ fixed to the values listed in Table~\ref{table:bestfitwlc}, but we still fitted for the quadratic limb darkening coefficients, independently for each channel. No temporal binning was applied to the spectroscopic lightcurves, since the RMS of the residuals behaves as expected for pure white noise after removing the common-mode signal (Figure~\ref{fig:allan}, bottom panel). For computational reasons, we decreased the number of live points to 1000 and the convergence tolerance to $\Delta\log Z < 0.01$. To assess the impact of limb-darkening assumptions, we produced an additional reduction at $\Delta\lambda = 0.004~\rm{\mu m}$ with quadratic limb-darkening coefficients fixed to the values from the \texttt{ExoTiC-LD} package \citep{grantwakeford2022exoticld}, based on 3D stellar models \citep{magic2015stagger} and the stellar parameters from \citet{hobson2023}: $T_{\rm eff}=5255$~K, $\log g = 4.582$, and $\lbrack \rm{Fe/H} \rbrack = 0.22$. In total, we produced three reductions with \texttt{Eureka!}:
\begin{enumerate}[label=(\alph*), itemsep=0pt, topsep=0pt]
\item $\Delta\lambda = 0.004~\rm{\mu m}$, with free limb darkening (our fiducial reduction).
\item $\Delta\lambda = 0.025~\rm{\mu m}$, with free limb darkening.
\item $\Delta\lambda = 0.004~\rm{\mu m}$, with fixed limb darkening.
\end{enumerate}

\subsection{\texttt{Tswift} reduction}\label{subsubsec:Second reduction}

We performed an additional independent reduction of the NIRSpec dataset with the Transit swift routine (\texttt{Tswift}). The general steps are similar to those used in \citet{fu2022}. The default JWST pipeline is first used to process the \texttt{uncal.fits} files to produce the \texttt{darkcurrentstep.fits} files. Then the spectral trace is masked out, and the unilluminated regions are used to perform column-by-column 1/f and background subtraction at the group level. Next, we use the RampFitStep from the JWST pipeline to create the \texttt{rampfitstep.fits} files. Then, we cleaned the bad pixels and extracted the spectral trace. Because the source is outside the slit and we are capturing the flux of the PSF wings, the spectral trace is broadened compared to typical observations. We decided to extract the spectra using a large aperture, with a width of 30 pixels along the spatial direction. Then each frame is summed in the vertical direction to create the white and spectroscopic lightcurves. 

\begin{figure}
\centering
\includegraphics[width=\linewidth]{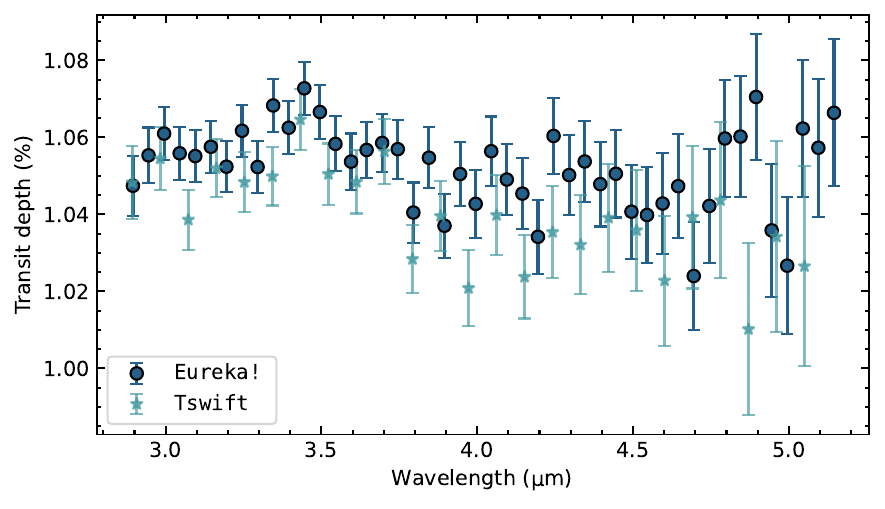}
\caption{Comparison of the two independent reductions of the NIRSpec data of TOI-199\,b.}\label{fig:comparison_reductions}
\end{figure}

We also measured the spectral trace drift as a function of time and, as the \texttt{Eureka!} reduction, found a sudden small jump in the \textit{y} direction around the 6340th integration, approximately 12.8 hours after the beginning of the observation (Figure~\ref{fig:white_lightcurve}). Due to the source being mostly outside the slit, this smaller drift in telescope pointing leads to significant wavelength-dependent slit-loss. To correct for this jump, we fitted a scaling factor for all points after the 6340th integration. So for the white lightcurve fit, we used \texttt{batman} \citep{kreidberg2015batman} with eight total free parameters, including a linear slope, constant, $R_p$, $a/R_s$, inclination, mid-transit time, y-shift coefficient, and a scale factor for all points after the 6340th integration. Then we fixed the best-fit white light $a/R_s$, inclination, and mid-transit time values for the spectroscopic lightcurve fits while fitting for the remaining five parameters. The limb darkening parameters are fixed to the 3D Stagger-grid stellar models \citep{magic2015stagger}. To check the robustness of fitting a scaling factor for all points after the 6340th integration, we also fit for the lightcurves with trimming all points after the 6340th integration. The resulting transit spectra from the two methods generally agree with a consistent shape. However, the trimming method results in larger error bars compared to the scaling factor approach due to the lack of egress and the subsequent baseline. The transit spectra from both methods are consistent with the one from \texttt{Eureka!}. In Figure~\ref{fig:comparison_reductions}, we compare the transmission spectra from the two independent reductions. Apart from a small vertical offset, the spectra show overall agreement.

\section{Atmospheric Retrievals}\label{sec:retrievals}
We performed Bayesian atmospheric retrievals to interpret the transmission spectra. To assess the robustness of our results, we applied two independent codes: \exotr\ \citep[][]{damiano2024} and Aurora \citep{welbanks2021a}.

\subsection{ExoTR}\label{subsubsec:exotr}
\exotr\footnote{https://github.com/MDamiano/ExoTR} is an atmospheric forward model and retrieval framework, which uses a nested sampling algorithm via \texttt{MultiNest} \citep{Feroz2009} and \texttt{PyMultiNest} \citep{Buchner2014} to statistically interpret exoplanet transmission spectra. \exotr\ maps molecular mixing ratios to their observable signatures, and it includes the ability to model realistic water clouds, hazes, and stellar activity \citep[][]{damiano2024}. The framework has been applied across diverse targets—LHS-1140\,b (cold, potentially water-rich), L\,98-59\,b (potentially volcanic), Kepler-51\,d (a hazy super-puff), and K2-18\,b (temperate sub-Neptune)—to infer atmospheric conditions from transmission data \citep{damiano2024, belloarufe2025, libbyrobertsarxiv, hu2025}.

\begin{table*}
\caption{Prior and Posterior Parameter Distributions from the \texttt{ExoTR} fit to the reference \texttt{Eureka!} reduction. For each species, we also report the Bayes Factor and corresponding Jeffreys interpretation, obtained by comparing the canonical model to a model excluding that parameter. }
\centering
\begin{tabular}{l c c c c}
\hline\hline
\textbf{Parameter}  & \textbf{Prior$^{(2)}$} & \textbf{Posterior} & \textbf{Bayes Factor} & \textbf{Jeffreys Scale$^{(3)}$} \\
\hline
Planetary radius [$R_p^{(1)}$] & $\mathcal{U}$(0.5, 2) & $0.97^{+0.01}_{-0.02}$ & - & -  \\
Atmospheric temperature [K] & $\mathcal{U}$(100, 500) & $325^{+88}_{-82}$ & - & - \\
Cloud top pressure [Pa]  & $\mathcal{LU}$(-1.0, 6.0) &  $2.3^{+1.4}_{-1.0}$ & 2.29 & Barely worth mentioning \\
VMR H$_2$O  & $\mathcal{LU}$(-12, -0.3) & $<-2.05$ & 1.07 & Barely worth mentioning \\
VMR CH$_4$  & $\mathcal{LU}$(-12, -0.3)  & $-2.46^{+0.85}_{-1.15}$ & 790.16 & Decisive evidence \\
VMR NH$_3$  & $\mathcal{LU}$(-12, -0.3) & $<-1.78$ & 1.26 &  Barely worth mentioning \\
VMR HCN  & $\mathcal{LU}$(-12, -0.3) & $<-1.94$ & 2.14 &  Barely worth mentioning \\
VMR CO & $\mathcal{LU}$(-12, -0.3) & $<-3.55$ & 0.95 & Barely worth mentioning (against)  \\
VMR CO$_2$  & $\mathcal{LU}$(-12, -0.3) & $<-3.47$ & 1.15 & Barely worth mentioning \\
VMR SO$_2$ & $\mathcal{LU}$(-12, -0.3)  & $<-4.85$ & 0.78 & Barely worth mentioning (against) \\
VMR OCS & $\mathcal{LU}$(-12, -0.3)  & $<-3.12$ & 2.75 & Barely worth mentioning\\
\hline
\end{tabular}
\vspace{5pt}
\tablecomments{$(^1)$ \citet{hobson2023}. $(^2)$ $\mathcal{U}(a,b)$ and $\mathcal{LU}(a,b)$ are the uniform and log-uniform (Jeffreys) distributions between values $a$ and $b$, respectively. Posteriors: Median values with 1$\sigma$ uncertainties. In the case of gases for which we did not obtain a detection, we present the upper limits corresponding to 95\% (2$\sigma$) cumulative probability levels.  $(^3)$ \citet{jeffreys1939}. Interpretations labeled ``against'' indicate that the data favor models excluding that parameter (i.e., Bayes factor $<1$).}
\label{tab:exotrpriors}
\end{table*}

We applied \exotr\ to interpret the spectrum of TOI-199\,b, fitting for a selected list of molecules and cloud parameters with uniform priors (Table~\ref{tab:exotrpriors}), and with H$_2$ as the fill gas. The selection of the molecules in our canonical model was guided by insights from self-consistent atmospheric models \citep[e.g.,][and those developed in this work]{hu2021photochemistry}. The molecular opacities were calculated line-by-line using the most up-to-date line lists from HITEMP \citep{Hargreaves2020} (CH$_4$), ExoMol \citep{Tennyson2016} (SO$_2$), and HITRAN \citep{Gordon2022} (all other molecules). Both the opacity grid and the forward-model evaluation use a resolution of $R=200{,}000$. Fixed parameters include the planet mass $M_p = 54.04$~M$_{\oplus}$ and stellar properties (e.g., photospheric temperature of $T_s=5255$K) \citep{hobson2023}.

\begin{figure*}
\centering
\includegraphics[width=1\linewidth]{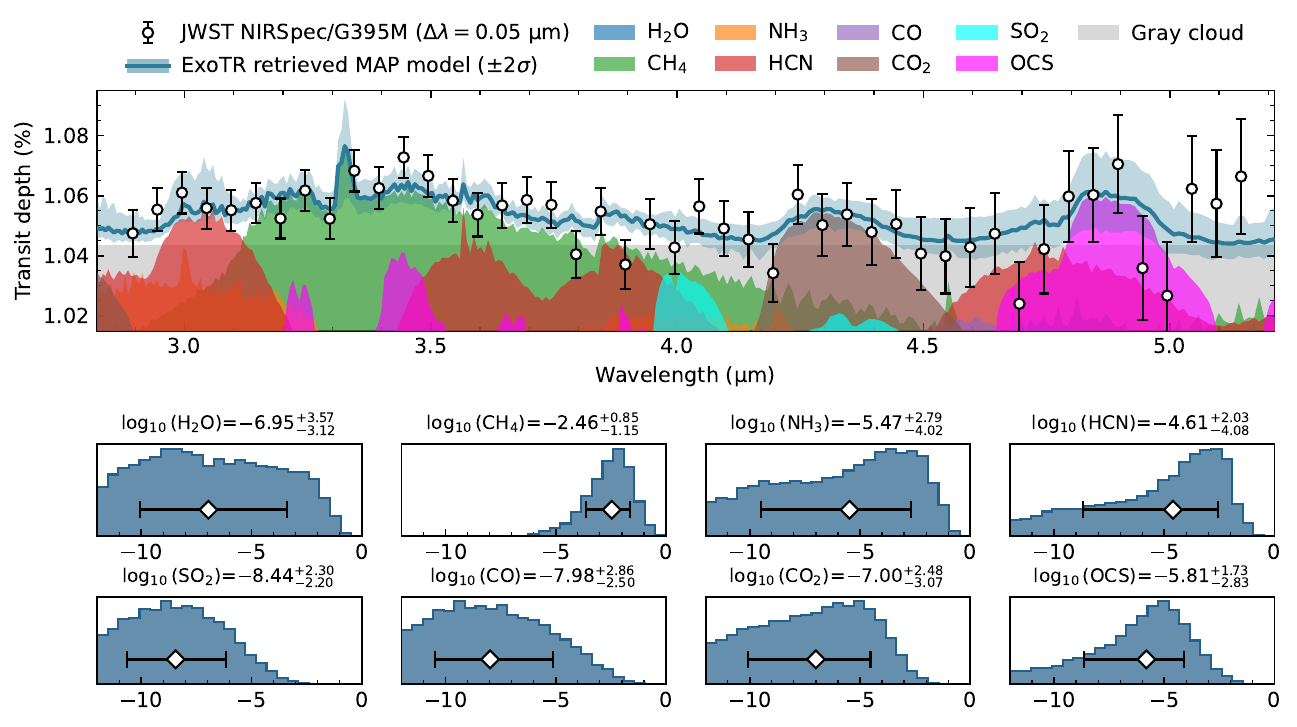}
\caption{Top: maximum \textit{a posteriori} (MAP) model and 2$\sigma$ credible regions from \exotr\ retrievals on the fiducial reduction, overlaid with data binned at $\Delta\lambda = 0.05~\mu\rm m$. Also shown are the contributions from each molecule and clouds to the transmission spectrum. Bottom: 1D histograms of the posterior distributions from the retrieval results, including the median and $\pm 1 \sigma $ uncertainties (i.e., 16$^{\rm th}$ and 84$^{\rm th}$ percentiles).}\label{fig:exotr_posteriors}
\end{figure*}

Figure~\ref{fig:exotr_posteriors} shows the best-fit model and molecular contributions from the retrieval. Many features are dominated by CH$_4$, especially in the 3.2-3.7~$\mu$m range; indeed, we are able to constrain CH$_4$ abundance to $\sim$1~dex and obtain a Bayes factor of 790, suggestive of a significant preference in favor of a model including CH$_4$ relative to a CH$_4$-free model. Other features are weakly fit with HCN, OCS, and/or CO$_2$ but these molecules, along with H$_2$O, NH$_3$, CO, SO$_2$, and H$_2$S, remain unconstrained. The 1D histogram for HCN and OCS suggests a hint of their presence, but the wide tail toward low abundance weakens any claim of robustness. Table~\ref{tab:exotrpriors} highlights the 1$\sigma$ confidence interval of the constrained species (CH$_4$) and the 2$\sigma$ upper bounds for the unconstrained molecules.

\begin{figure*}
\centering
\includegraphics[width=1.\linewidth]{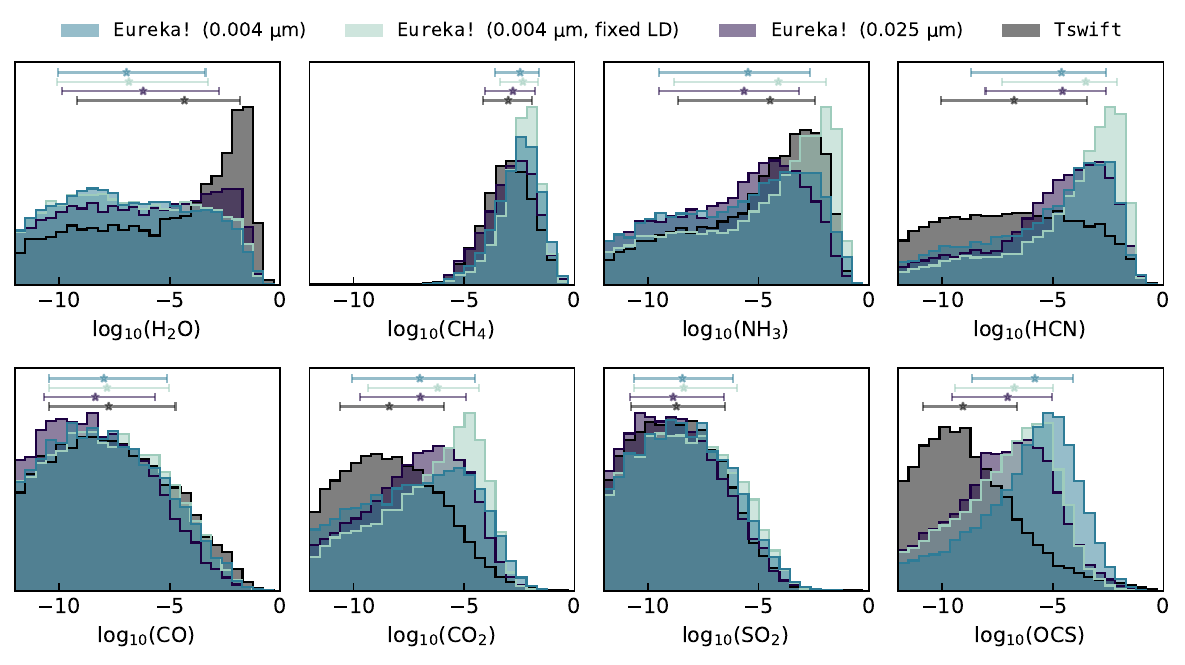}
\caption{Posterior distributions of the gas abundances from the \texttt{ExoTR} retrievals on the different data reductions. Medians and 1$\sigma$ uncertainties are overplotted as markers with horizontal error bars.}\label{fig:exotr_posteriors_comparison}
\end{figure*}

We repeated the retrieval for three additional cases: the native spectral resolution data with fixed limb-darkening coefficients, data binned down to a resolution of 0.025~$\mu$m, and the data resulting from the \texttt{Tswift} reduction. Slight differences arise in the posterior solutions, but the main conclusions remain the same (see Figure~\ref{fig:exotr_posteriors_comparison}).

\begin{figure*}
\centering
\includegraphics[width=\linewidth]{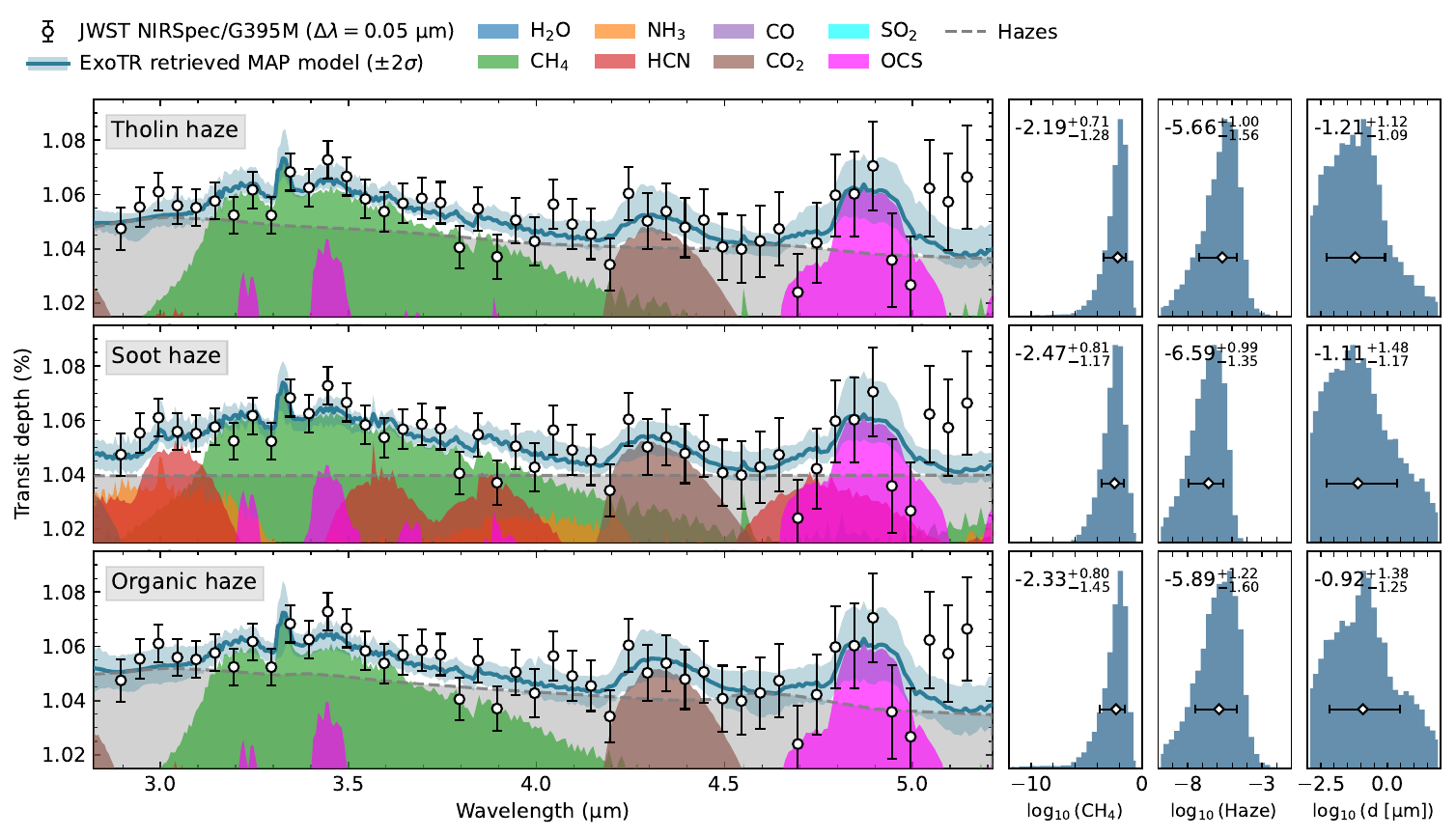}
\caption{Same as Figure~\ref{fig:exotr_posteriors}, but for models that include a tholin haze (top), a soot haze (middle), and an organic haze (bottom). Next to each spectrum we show the corresponding 1D histograms of methane and the haze abundance and particle diameter.}\label{fig:exotr_posteriors_haze}
\end{figure*}

We explored the effect of aerosols by repeating the baseline retrieval without the gray cloud, considering both a cloud/haze-free (clear) atmosphere and scenarios with a Mie-scattering haze, using optical properties of tholin \citep[i.e., Titan haze analog,][]{khare1984}, soot \citep[i.e., absorbing black carbon,][]{Hess1998}, and organic haze \citep[i.e., tholin-like aerosols experimentally produced in 400-K water-rich environments,][]{He2024} individually to assess both the type and properties of hazes constrained by the data (see Figure~\ref{fig:exotr_posteriors_haze}).
We found no significant preference between a cloudy and cloud-free model, with a Bayes factor of only 2.29 for clouds. The Bayes factors for hazy atmosphere models with tholin, soot, or organics are similarly small, suggesting ambiguity in the data for the presence of haze (Table~\ref{tab:exotrhaze}). However, given the relationship between the atmospheric CH$_4$ and haze production, particularly for tholin and organics, we also test whether the evidence for haze increases for a CH$_4$-free model. Indeed, we find that the Bayes factor does not change for soot, which is simple carbon, but does increase significantly for tholin and organics (up to 35 and 36, respectively, or ``very strong" evidence according to \citealt{thorngren2026}). Thus, the evidence for CH$_4$ in a hazy atmosphere correspondingly decreases, from 291 to 22 and 16 for tholin and organics, respectively (and up slightly to 471 for soot). The haze particle diameters in these models are quite small (median of 0.015~$\mu$m) and the abundances are quite high (median of 40~ppm), suggesting that an unusually large cross section for haze is needed to make up for the lack of CH$_4$. Also, when comparing model evidence for CH$_4$ vs tholin/organic haze in analogous cases, there is always a preference for the CH$_4$ model by a factor of $\sim 8$. Since the presence of CH$_4$ is directly related to tholin and organic hazes, whether one model is preferred over the other does not change our main conclusion that there is strong evidence for CH$_4$ in the atmosphere of TOI-199\,b. We also note that the mixing ratio of \ce{CH4} remains nearly constant across all haze and gray cloud models, with a maximum difference in the median abundances of $\Delta\log(\rm{CH}_4)=0.28$, well below the 1$\sigma$ uncertainties. To more rigorously constrain the presence or absence of clouds and hazes, additional transit data, especially at shorter wavelengths where a haze slope would be more prominent, would be especially helpful and would help to distinguish between a cloudy, clear, and hazy atmosphere.

\begin{table}[]
\caption{Bayes factor (BF) for CH$_4$ and various haze types from \exotr\ retrievals.}
\centering
\begin{tabular}{@{}lccc@{}}
\hline\hline
Baseline Case & BF(\ce{CH4}) & BF(Haze) & Ratio \\
\hline
\multirow{3}{*}{clear, no CH$_4$} 
 & \multirow{3}{*}{291} & 35 (tholin)   & 8.3 \\
 &                      & 36 (organics) & 8.1 \\
 &                      & 1.6 (soot)    & 180 \\
\hline
tholin w/ CH$_4$     & 22.4 & 2.68 & 8.4 \\
\hline
organics w/ CH$_4$   & 16.1 & 2.00 & 8.1 \\
\hline
soot w/ CH$_4$       & 471 & 2.61 & 180 \\
\hline
\end{tabular}
\label{tab:exotrhaze}
\tablecomments{For the first block, the baseline is a clear atmosphere with no CH$_4$. We report the Bayes factor for adding CH$_4$ vs.\ adding each haze type individually. For the remaining three rows, the baseline contains both haze and CH$_4$, and we assess the Bayes factor by removing either the CH$_4$ or the haze, one at a time.}
\end{table}

\subsection{Aurora}\label{subsubsec:aurora}

Aurora \citep{welbanks2021a} is a Bayesian retrieval and forward-modeling framework for interpreting exoplanet transmission and emission spectra \citep[e.g.,][]{Bell2023, Welbanks2024}. It solves the radiative transfer equation in a plane-parallel geometry under hydrostatic equilibrium, with flexible parameterizations of the atmospheric temperature structure, composition, and cloud/haze properties. Using this complementary framework, we tested the dependence of the retrieved atmospheric properties on the specific model assumptions considered in the main analysis from \exotr\ in Section \ref{subsubsec:exotr}. This approach allows us to assess how model degeneracies \citep[e.g.,][]{Welbanks2019a,Welbanks2022} and model construction choices \citep[e.g.,][]{welbanks2026} affect the interpretation of the atmospheric spectrum. Parameter estimation was performed using \texttt{MultiNest} \citep{Feroz2009} via \texttt{PyMultiNest} \citep{Buchner2014}.

We explored three sets of modifications relative to the \exotr\ baseline model to evaluate the robustness of retrieved CH$_4$, NH$_3$, and HCN abundances. Specifically, we considered 1) the relaxation of the isothermal temperature assumption, and 2) incorporating inhomogeneous cloud and haze cover. We performed retrievals using the model configurations described below on all reductions. 

These atmospheric models considered the same absorbers as \exotr\ (e.g.,
H$_2$O \citep{Rothman2010}, CH$_4$ \citep{Yurchenko2014a}, NH$_3$ \citep{Yurchenko2011a}, HCN \citep{Barber2014a}, CO \citep{Rothman2010}, CO$_2$ \citep{Rothman2010}, SO$_2$ \citep{Underwood2016a}, OCS \citep{Wilzewski2016}). The opacities were computed line-by-line from the HITRAN \citep{Rothman2010, Richard2012}, HITEMP \citep{Hargreaves2020}, and ExoMol \citep{Tennyson2016} databases, following the prescriptions in \citet{welbanks2021a, Welbanks2024}. 

Aurora accounts for the effects of inhomogeneous clouds and hazes modeled as a linear combination of a clear and cloudy/hazy region, following the cloud deck prescription and enhanced H$_2$-Rayleigh scattering described in \citet{welbanks2021a}, which builds upon the works of \citet{Etangs2008a, Line2016a, MacDonald2017}. For the non-isothermal temperature structure, we adopted the six parameter prescription from \citet{Madhusudhan2009}.  All model spectra were computed at a resolving power of $R=50,000$ over the wavelength range of 2.5--5.5 $\mu$m.

\begin{figure*}
\centering
\includegraphics[width=.95\linewidth]{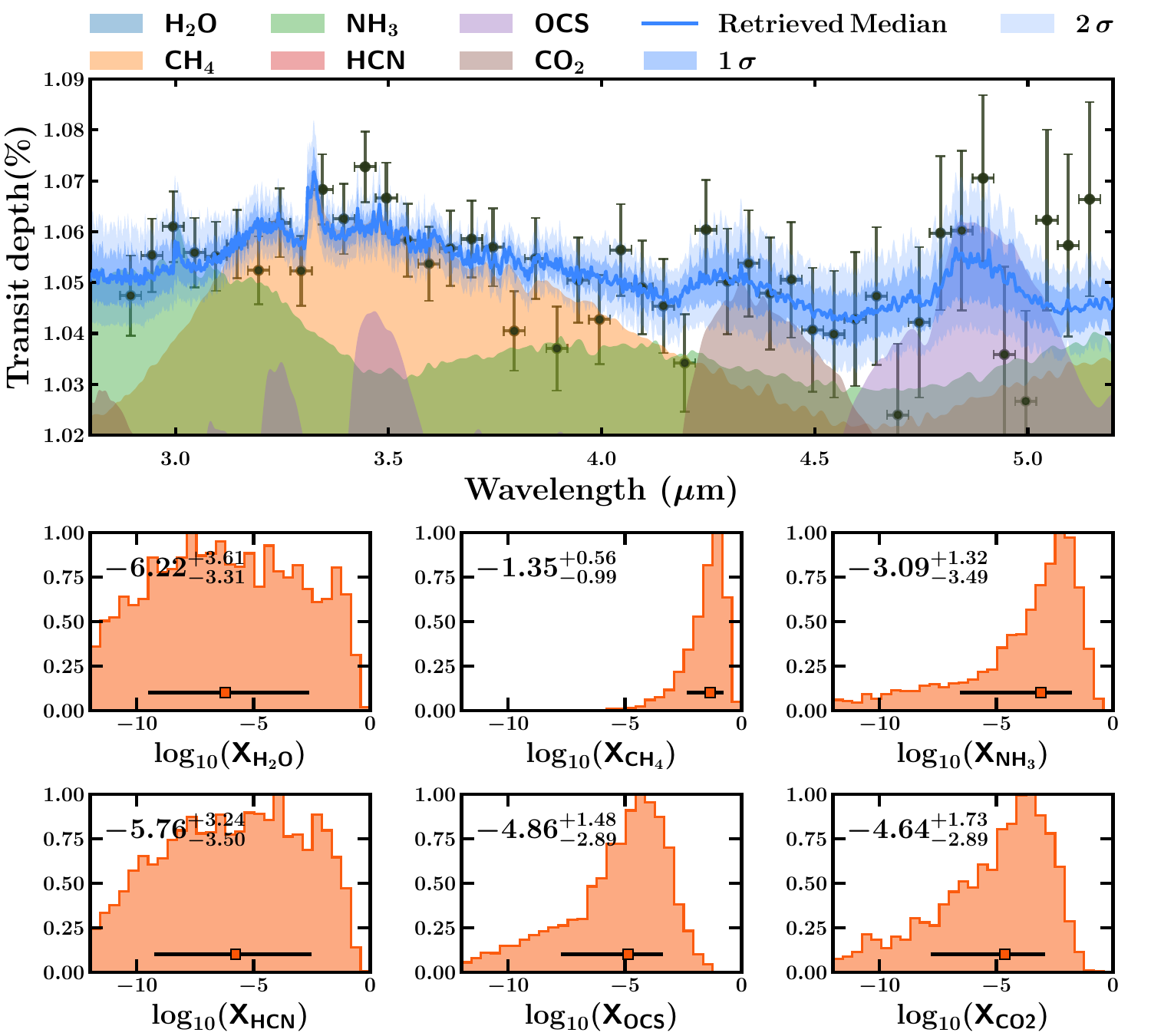}
\caption{Top: best-fit spectrum with 1$\sigma$ and 2$\sigma$ model posteriors, from retrievals using Aurora and molecule contributions from the best fit model. Bottom: Posterior distributions for the molecules of interest, including the median and $\pm 1 \sigma$ uncertainties.}\label{fig:aurora_posteriors}
\end{figure*}

The retrieved spectra and posteriors from the analysis on the \texttt{Eureka!} reduction with $\Delta \lambda=0.004$ $\mu$m are shown in Figure \ref{fig:aurora_posteriors}. For visual clarity, the shown data are binned. The retrieved atmospheric temperature structure is non-isothermal with  T=$494^{+283}_{-157}$~K at 100~mbar, which is consistent with the planet's equilibrium temperature at the 1-$\sigma$ level. The retrieved abundances of NH$_3$, H$_2$O, CO, CO$_2$, HCN, and OCS remain poorly constrained as with the analysis from \exotr. However, the spectrum near $\sim3$ $\mu$m seems to be preferentially explained by the absorption line of NH$_3$ rather than clouds and HCN as with the \exotr\ analysis. A model comparison between the model including the species and the model excluding it suggests no meaningful preference for the inclusion of NH$_3$.

Nonetheless, the interpretation of the spectrum of TOI-199\,b as being largely explained by the methane feature at $\sim3.3\mu$m and its broadband feature remains consistent. The retrieved CH$_4$ abundance is comparable with that derived from \exotr\ although generally higher. This is consistent with expectations from known degeneracies between molecular abundances and the presence of inhomogeneous clouds/hazes \citep[e.g.,][]{Welbanks2019a, welbanks2021a, Nixon2024a}. The preference for the model including CH$_4$ over the model without the absorber is marginally lower, with a Bayes factor of 705. The preference for the full over a model without NH$_3$ or without NH$_3$ and HCN is weak, with Bayes factors of 2. We emphasize that this model preference is relative to the specific set of molecules considered.
Furthermore, including more species such as C$_2$H$_6$, C$_2$H$_2$, and C$_2$H$_4$, or other molecules considered in the full forward models (presented in Section \ref{sec:self-consistent}), could alter the outcome \citep{welbanks2026}.

\section{Self-Consistent Photochemistry Models}\label{sec:self-consistent}
For a more in-depth exploration of the data, we now proceed to compare the NIRSpec G395M transmission spectrum of TOI-199\,b against a series of self-consistent photochemical models at different metallicities, C/O ratios, internal temperatures, and vertical mixing efficiencies ($K_{\rm zz}$). The first goal of this exercise is to determine whether the current data can place meaningful constraints on any of these parameters and evaluate the degree of agreement between the retrieval results and self-consistent photochemical models in radiative-convective equilibrium. The second goal is to identify the spectral regions where these parameters will make an appreciable difference and help identify future observation opportunities for \planetname.






\subsection{Model setup}


We modeled the atmosphere of TOI-199\,b using the climate module of \texttt{EPACRIS} \citep{hu2024}, previously applied in \citet{damiano2024,Yang2024_epacris,belloarufe2025}. We explored both solar and enhanced metallicity cases ($\rm{M/H} = 10$) and considered solar (0.59) and super-solar (1.1) C/O ratios. While not shown, we find that increasing the C/O ratio to 1.1 has only a weak effect on the transmission spectrum. Because gas giants span a wide range of possible internal temperatures depending on mass, age, evolutionary history, and orbital eccentricity \citep{fortney2020beyond}, we examined models ranging from a nominal internal temperature of $T_{\rm int} = 50$~K up to 150 K. To explore the role of vertical mixing, we explored several eddy diffusion coefficient ($K_{\rm zz}$) profiles (Figure~\ref{fig:TP_KZZ}), including the nominal profile with a $K_{\rm zz}$ of $10^{6}\ \mathrm{cm}^2\,\mathrm{s}^{-1}$ in the convective region and $10^{8}\ \mathrm{cm}^2\,\mathrm{s}^{-1}$ in the deep atmosphere \citep{zhang2018global1}, and a Jupiter-like profile with much weaker mixing in the convective region \citep[$K_{\rm zz}\sim10^{3}\ \mathrm{cm}^2\,\mathrm{s}^{-1}$,][]{gladstone1996hydrocarbon}. We also explored uniform $K_{\rm zz}$ spanning $10^{4}$–$10^{8}\ \mathrm{cm}^2\,\mathrm{s}^{-1}$ as assumed by most hot-Jupiter studies, but we did not find significant deviations from the more realistic models.
Further details of the self-consistent model setup are provided in Appendix~\ref{app:models}.

\subsection{Self-consistent model results}
\begin{figure*}[t]
    \centering
    \begin{subfigure}{0.32\textwidth}
        \centering
        \includegraphics[width=\linewidth]{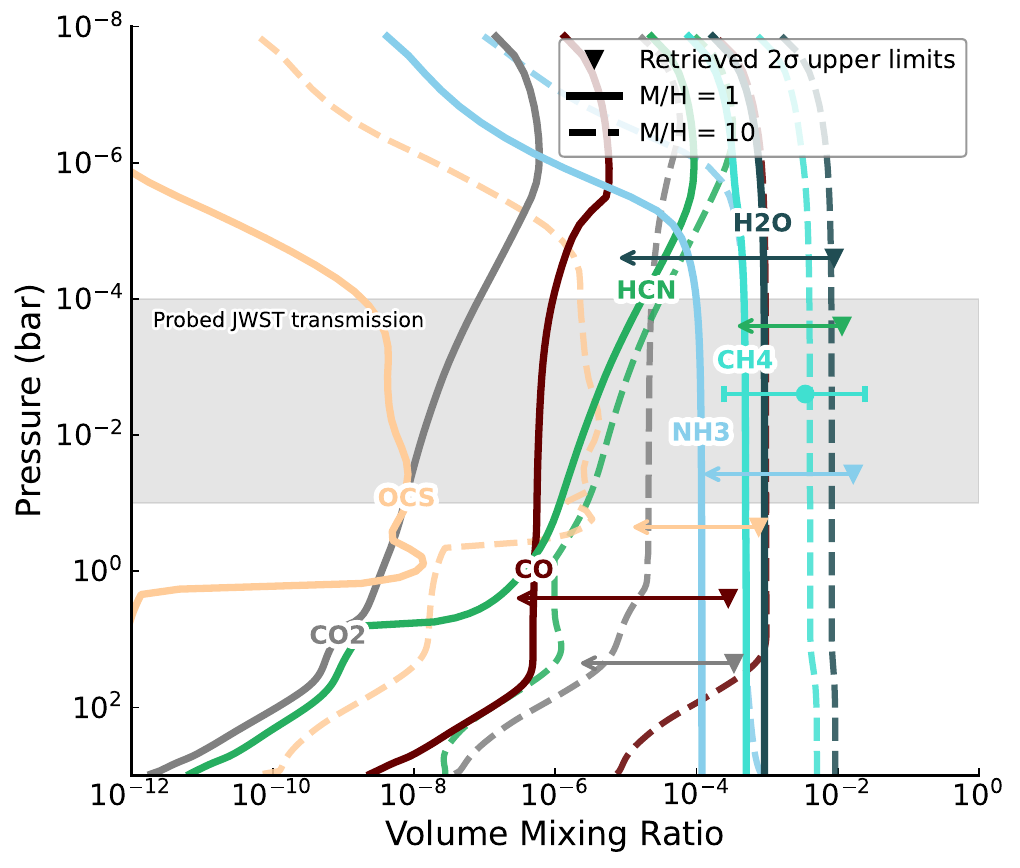}
        \caption{Metallicity}
        \label{fig:metal_comp}
    \end{subfigure}
    \hfill
    \begin{subfigure}{0.32\textwidth}
        \centering
        \includegraphics[width=\linewidth]{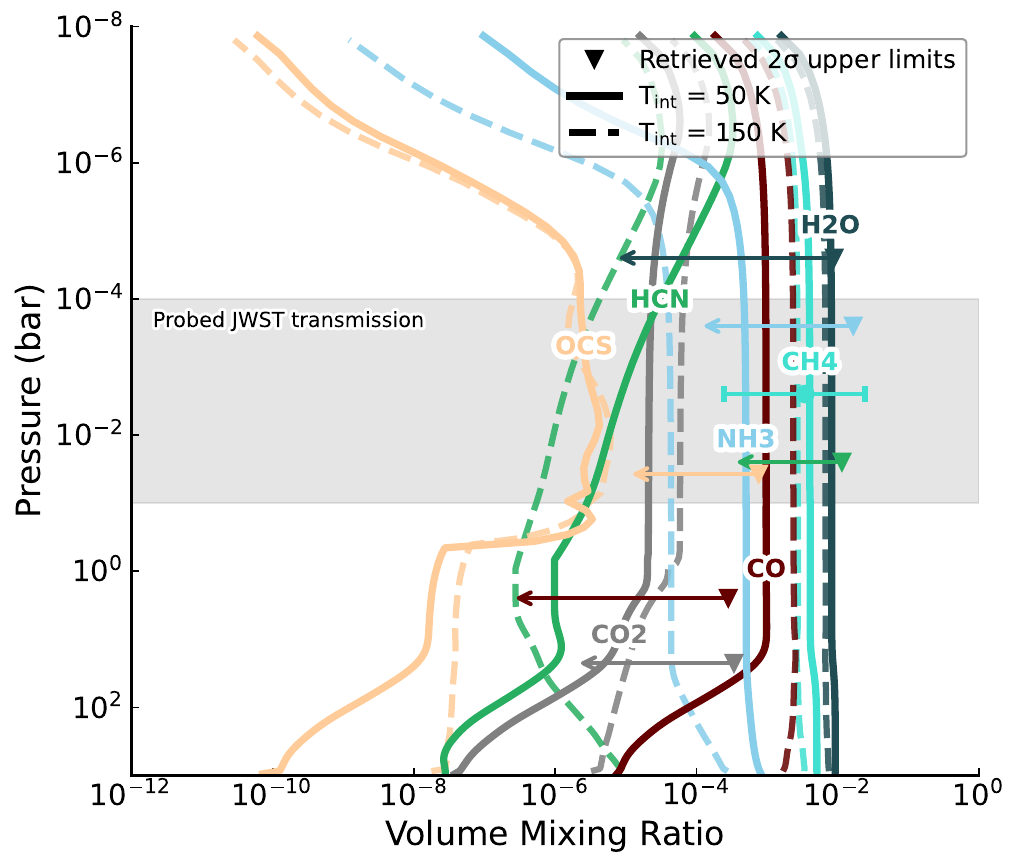}
        \caption{Internal temperature}
            \label{fig:main_chem}
    \end{subfigure}
    \hfill
    \begin{subfigure}{0.32\textwidth}
        \centering
        \includegraphics[width=\linewidth]{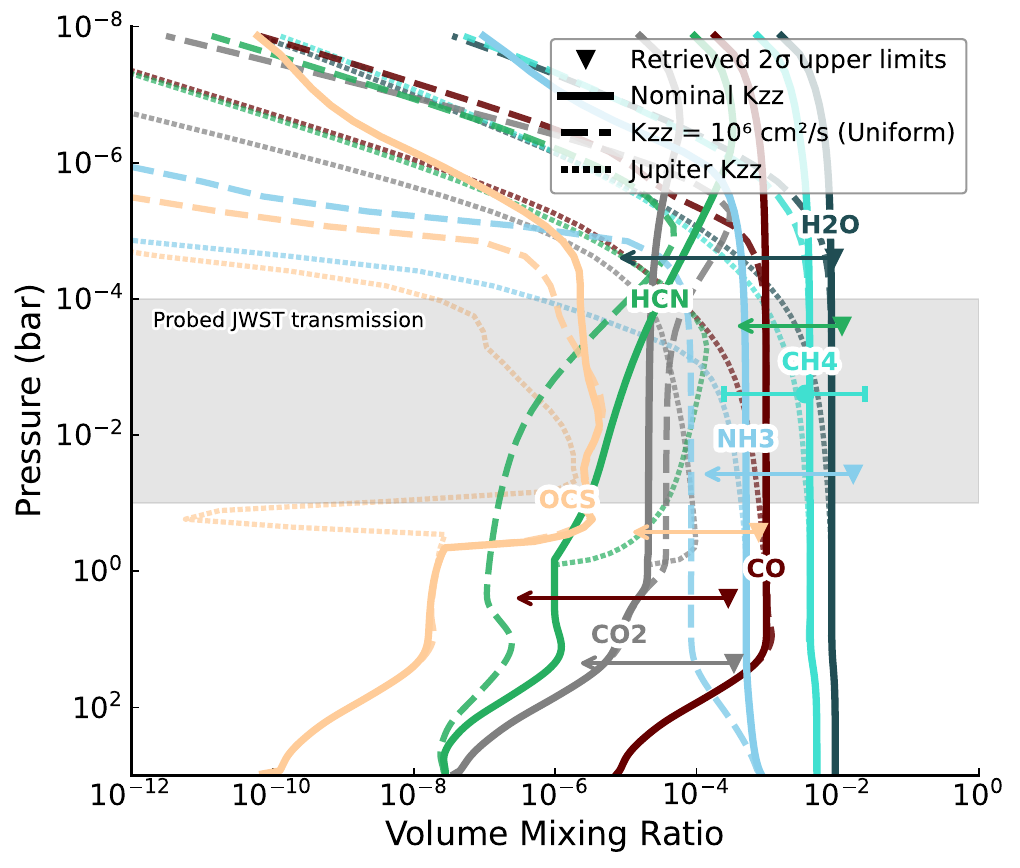}
        \caption{Eddy diffusion}
        \label{fig:kzz_comp}
    \end{subfigure}
    \caption{Effects of key atmospheric parameters on the photochemical equilibrium abundances in \planetname. 
    \textbf{(a)} Comparison between solar and enhanced metallicity atmospheres, assuming a solar C/O ratio and $T_{\rm int} = 50\,\rm K$. Colored arrows denote the retrieved \ce{CH4} abundance ($1\sigma$) from \texttt{ExoTR} and $2\sigma$ upper limits for other species, with arrow lengths indicating $1\sigma$ uncertainties. The gray shaded region marks the approximate pressure range probed in transmission.
    \textbf{(b)} Comparison between models with $T_{\rm int} = 50\,\rm K$ and $T_{\rm int} = 150\,\rm K$, assuming enhanced metallicity ($\rm M/H = 10$).
    \textbf{(c)} Comparison of various eddy diffusion ($K_{\rm zz}$) profiles, assuming $\rm M/H = 10$ and $T_{\rm int} = 50\,\rm K$.}
    \label{fig:}
\end{figure*}

Figure \ref{fig:metal_comp} compares solar and enhanced metallicity atmospheres. The main components (i.e., \ce{CH4}, \ce{NH3}, and \ce{H2O}) scale nearly linearly with metallicity. \ce{CH4} and \ce{NH3} remain within the observed uncertainty range at both solar and $10\times$solar metallicity.
While the deep-atmosphere abundance of \ce{HCN} scales sharply with metallicity, its upper-atmosphere abundance is less affected as it is also governed by the balance between photochemical loss and vertical mixing \citep{hu2021photochemistry}. The atmospheric abundance of \ce{CO} and \ce{CO2} scales nonlinearly with the metallicity, and metallicity values larger than $\rm{M/H} = 10$ push \ce{CO} and \ce{CO2} beyond the observed $2\sigma$ upper limits, making significantly higher metallicities unlikely for \planetname.

Figure~\ref{fig:main_chem} shows a comparison of major photochemical abundances for an atmosphere with enhanced solar metallicity ($\rm{M/H} = 10$) and nominal $K_{\rm zz}$ for internal temperatures of  $T_{\rm int} = 50\ \rm K$ and $T_{\rm int} = 150\ \rm K$. Increasing the internal temperature has important implications for the resulting composition: while \ce{CH4} and \ce{H2O} remain the main carbon and oxygen carriers, \ce{CO} increases substantially and nearly overtakes \ce{CH4}. Higher temperatures in the deep atmosphere also suppress \ce{NH3}, decreasing its VMR by approximately an order of magnitude. This also causes the abundance of \ce{HCN} in the observable part of the atmosphere to decrease substantially, because \ce{HCN} is derived photochemically from 
\ce{NH3}.

Figure \ref{fig:kzz_comp} illustrates the impact of vertical mixing on atmospheric composition. For \ce{NH3}, a strong deep-atmosphere mixing ($K_{\rm zz} = 10^8\ \mathrm{cm}^2\,\mathrm{s}^{-1}$), which is assumed in the nominal and Jupiter-like cases, leads to quenching at higher pressures and results in atmospheric abundances reflective of the overall nitrogen abundance of the envelope. However, if $K_{\rm zz}$ is reduced to $10^6\ \mathrm{cm}^2\,\mathrm{s}^{-1}$ in the deep atmosphere, \ce{NH3} is quenched at $\sim10$ bar, resulting in a VMR decrease by a factor of 10 with respect to the $K_{\rm zz} = 10^8\ \mathrm{cm}^2\,\mathrm{s}^{-1}$ cases. Consequently, the abundance of \ce{HCN} in the observable part of the atmosphere is also suppressed. 

Between the nominal and Jupiter-like cases, the eddy diffusion coefficient $K_{\rm zz}$ in the convective region differs significantly, leading to substantial differences in the amount of \ce{HCN} that accumulates in the observable atmosphere. Although \ce{HCN} originates from similar \ce{NH3} abundances, a bottleneck in vertical transport -- such as in the Jupiter-like case, where $K_{\rm zz}$ is small in the upper troposphere -- allows photochemically produced \ce{HCN} to accumulate more efficiently in the upper atmosphere, resulting in a higher steady-state abundance (Figure~\ref{fig:kzz_comp}). Consequently, the \ce{HCN}/\ce{NH3} ratio in the observable atmosphere provides a sensitive diagnostic of the efficiency of vertical transport in the atmosphere of \planetname.

\subsection{Comparison to the observations}

\begin{figure*}
    \centering
    \includegraphics[width=1.\linewidth]{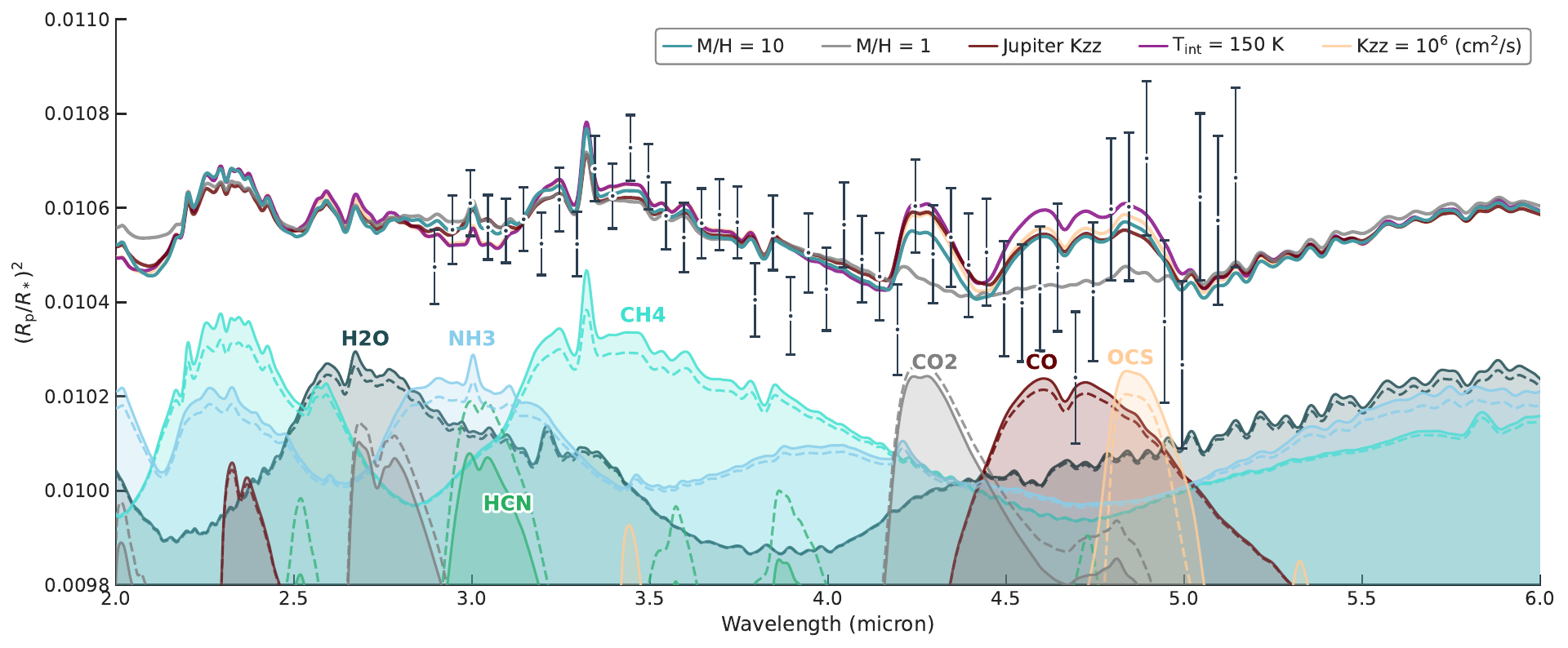}
\caption{Self-consistent model transmission spectra comparison for TOI-199\,b. The teal spectrum represents an enhanced metallicity case ($\rm{M/H} = 10$) using the nominal $K_{\rm zz}$ profile. For enhanced metallicity, we also show spectra with a Jupiter-like $K_{\rm zz}$ profile (red), increased internal temperature (purple), and a uniform $K_{\rm zz}$ value of $10^6$~cm$^2$~s$^{-1}$ (orange). Additionally, the solar metallicity spectrum is shown in gray. Individual opacity contributions of the M/H = 10 model are shown via solid shaded regions. Individual contributions, shown in dashed lines, represent the Jupiter-like case. All spectra are offset to minimize $\chi^2$. NIRSpec data are shown for the fiducial \texttt{Eureka!} reduction binned at $\lambda/\Delta\lambda = 100$ for visualization.}
    \label{fig:model_comparison}
\end{figure*}

Figure \ref{fig:model_comparison} shows the resulting transmission spectra from our self-consistent models, comparing several cases that highlight differences in spectral features caused by metallicity, internal heating, and vertical mixing efficiency. The best-fitting models from our set are the enhanced-metallicity cases ($\rm{M/H} = 10$) with the Jupiter-like $K_{\rm zz}$ and a uniform $K_{\rm zz} = 10^{8}\ \mathrm{cm}^2\,\mathrm{s}^{-1}$ profile, both with a reduced chi-squared of $\chi^2_\nu =1.0469$. However, due to the lower-than-expected precision in our program, all of the presented models can reasonably explain the observed transmission spectrum. 

\section{TTV Analysis}\label{sec:ttvs}
In order to refine the physical and orbital properties of the two known planets in the TOI-199 system and predict future transit times, we combined the JWST data with TESS observations to perform an updated TTV analysis. Planetary mass affects the transmission spectrum through the atmospheric scale height; however, as described below, the derived mass for \planetname\ remains identical to that reported by \citet{hobson2023}. Consequently, the assumed scale height is unchanged and the atmospheric inferences are unaffected.

\begin{figure}
    \centering
    \includegraphics[width=1.\linewidth]{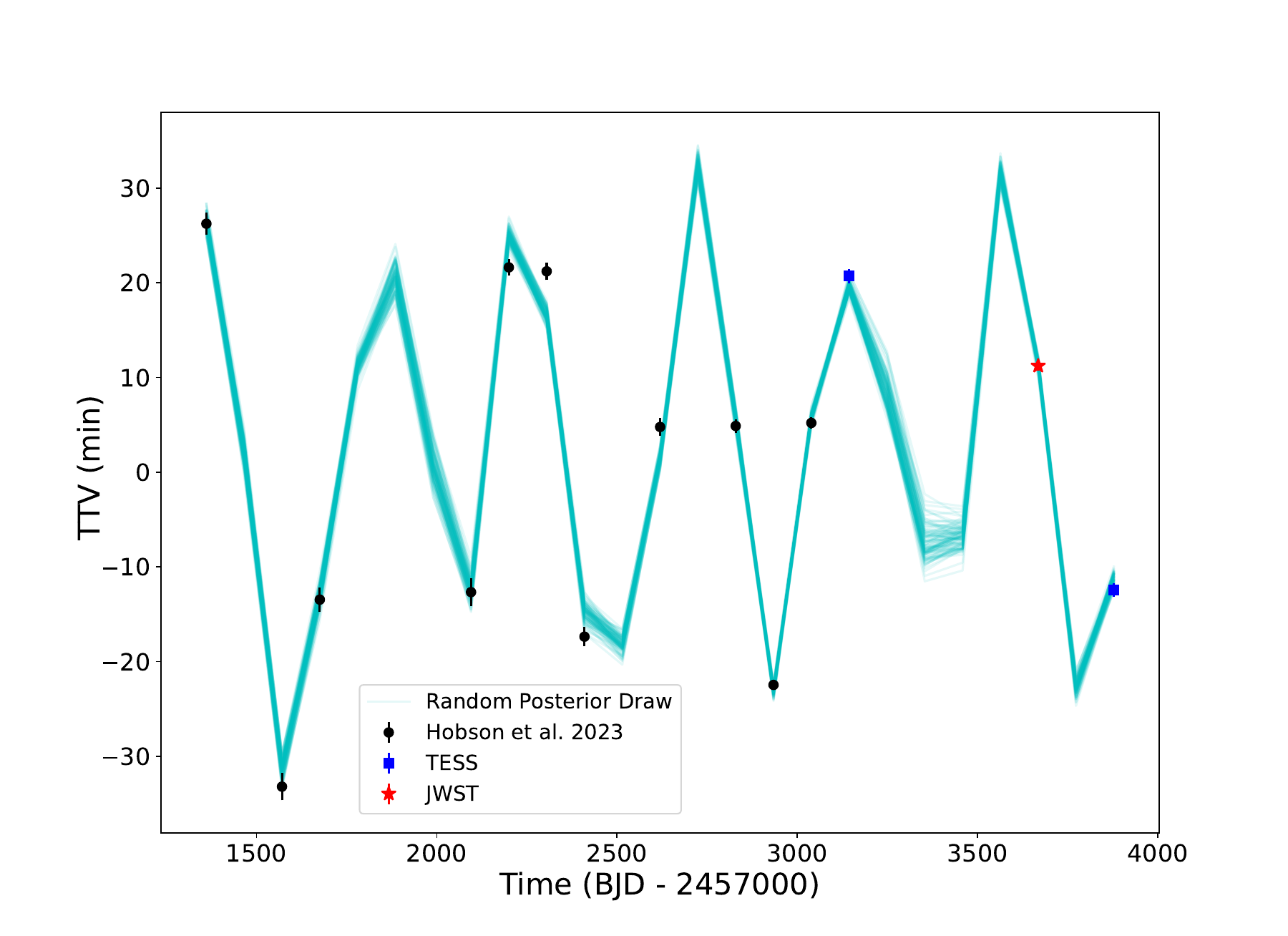}
\caption{Observed TTVs for TOI-199\,b from the \cite{hobson2023} discovery paper (black points) and new data from JWST (red star) and TESS sectors 67, 87, and 94 (blue squares), along with 100 random draws from the best-fit TTV model (blue lines).}
    \label{fig:TTVs}
\end{figure}

The strong TTVs in this system were first reported by \cite{hobson2023}, who performed a joint analysis between TTV and radial velocity data (RV) obtained by TESS and several ground-based observatories. This TTV+RV analysis revealed TTVs with a peak-to-peak amplitude around 60 minutes for TOI-199\,b and suggested a non-transiting planet with an orbital period of $\sim$274 days and mass of $\sim0.28\,M_J$, placing it in the conservative habitable zone. The TOI-199 system is relatively uncommon in that it exhibits strong TTVs, but the two known planets are not near any first- or second-order orbital resonances, which are typically responsible for causing observed TTVs \citep[e.g.,][]{Lithwick_2012,Holczer2016}. In the case of TOI-199, the two massive planets on eccentric and close-in ($<$ 1 AU) orbits dynamically perturb each other enough to cause large TTVs without proximity to resonance, which leads to TTVs that do not exhibit the typical sinusoidal pattern \citep[e.g.,][see also Figure \ref{fig:TTVs}]{Ford2012,Deck2016}. 

Since the original publication of TTVs in the TOI-199 system in \cite{hobson2023}, three more transits of TOI-199\,b have been observed in subsequent TESS sectors, including one additional transit prior to our scheduled JWST transit observation.
We extracted transit times for TOI-199\,b from TESS sectors 67, 87, and 94, using the same procedure as described in \cite{Greklek-McKeon2025a}. In brief, this involved constructing a TESS photometric model using the \texttt{exoplanet} package \citep{exoplanet:foremanmackey18} for the transit lightcurve component, and a Matern-3/2 Gaussian Process (GP) kernel from \texttt{celerite2} \citep{exoplanet:foremanmackey17} to model out-of-transit stellar variability. Our fits to the TESS data used 2-minute cadence data from the Science Processing Operations Center pipeline \citep[SPOC;][]{Jenkins2016}, which excluded $\sim$30\% of data that was flagged by the SPOC pipeline for poor data quality, including a potential transit event in sector 90. A more careful photometric extraction from a custom TESS aperture, similar to the technique of \cite{hobson2023}, may be able to recover an additional transit event for TOI-199\,b in future work. Our transit model includes parameters for the stellar radius $R_{\star}$, impact parameter b, scaled semimajor axis a/$R_{\star}$, planet-to-star radius ratio $R_p$/$R_{\star}$, and individual transit mid-center times to account for TTVs. The GP model includes a mean offset parameter $\mu$ and hyperparameters $\sigma$ and $\rho$ corresponding to the amplitude and timescale of quasiperiodic oscillations, along with an error scaling term added in quadrature to the TESS flux errors. We fixed the planetary eccentricities to zero in the transit fit, as in \cite{hobson2023}, since the corresponding effect on the transit lightcurve shape is negligible. We also fixed the quadratic limb-darkening parameters $u_1$ = 0.466 and $u_2$ = 0.117, as predicted by the \texttt{ldtk} package \citep{Parviainen_2015} using the stellar temperature, [Fe/H], and log g
values reported in \cite{hobson2023}. We placed Gaussian priors on $R_{\star}$, b, a/$R_{\star}$, and $R_p$/$R_{\star}$ using the values reported in Tables 2 and 3 of \cite{hobson2023}, and placed a uniform prior on individual transit times $\pm$ 12 hours from the transit time predicted from the \cite{hobson2023} ephemeris. We used the \texttt{PyMC3} package \citep{exoplanet:pymc3} to sample the posterior distribution of our model with four parallel chains run with 2000 burn-in steps and 2000 posterior sample draws. We confirmed that the chains evolved until the Gelman–Rubin statistic values were $<$1.01 for all parameters, and that the transit mid-time posteriors were unimodal and normally distributed. The transit times from sectors 67, 87, and 94, along with the best-fit JWST transit time, and predicted transit times through 2040 December 31, are listed in Table \ref{tab:transit_times}.

Using the new TOI-199\,b transit times, we performed an updated TTV fit following the procedure of \cite{Greklek-McKeon2025a}. This involved using the TTVFast package \citep{Deck2016} to model the observed transit times listed in Table \ref{tab:transit_times}. The modeled
transit times are a function of the planetary masses and orbital
elements relative to a reference epoch at the TTV model start time, which we chose to be 1256.14 (BJD$–$2\,457\,000) to match the simulation start time of \cite{hobson2023}. This simulation start time is close to one orbital period before the first observed transit of TOI-199\,b, which can make TTV model convergence more difficult due to slight changes in the orbital period parameter for TOI-199\,b altering the total number of transit events within the simulation time period, but we use this start time regardless for consistency with the \cite{hobson2023} model. 

In our TTV modeling, we fixed the planetary orbital inclinations to 90$^{\circ}$ and the stellar mass to 0.936 $M_{\odot}$, as in the original TTV analysis of \cite{hobson2023}. Our TTV model has 10 free parameters in total, including: the planet-to-star mass ratios, average Keplerian orbital periods over the simulation period, initial mean anomalies ($M_0$) re-parameterized as the time of first transit ($T_0$), and the planetary eccentricities ($e$) and longitudes of periastron ($\omega$). We re-parameterized the latter two quantities as $\sqrt{e}\cos(\omega)$ and $\sqrt{e}\sin(\omega)$ to mitigate the degeneracy between $e$ and $\omega$ in our fits while retaining an effective uniform prior on $e$, as recommended by \citep{exofast}. The orbital periods, mean anomalies, eccentricities, and longitudes of periastron are osculating orbital elements defined at the TTV model start time. We fit this model to the data using the MCMC ensemble sampler \texttt{emcee} \citep{foremanmackey2013emcee}, and chose wide
uniform priors for all parameters (see Table \ref{table:ttv_fit_results}) except the mass of TOI-199\,b, for which we use a Gaussian prior derived from the independent RV mass constraint of \cite{hobson2023}, which is scaled from the RV semi-amplitude constraint. 

We performed our TTV fit including the original transit times published in \cite{hobson2023}, the JWST transit time, and additional new TESS data from sectors 67, 87, and 94. The same TOI-199\,b transit observed by JWST was also observed by TESS in sector 87, so we used the error-weighted average of these transit times along with its corresponding uncertainty in the TTV model. We initialized this TTV fit with 300 walkers (30 per free parameter) with parameters initialized from the best-fit solution of \cite{hobson2023}, and ran the MCMC for 50,000 steps, discarding the first 20,000 steps as burn-in and ensuring convergence by verifying that each parameter reached more than 50 autocorrelation lengths in the MCMC.  

Our updated TTV model including the new JWST data and three additional TESS transits produced results for the masses of TOI-199\,b and c consistent within 2$\sigma$ with the results of \cite{hobson2023}. Our complete TTV model results are listed in Table \ref{table:ttv_fit_results}. Our derived mass for \planetname\ is identical to that reported by \citet{hobson2023}, and therefore our atmospheric inferences are unaffected. Meanwhile, our new results improve the precision on the mass of TOI-199\,c by approximately 50\%, and reduce the uncertainty in the eccentricity by more than a factor of 3 for both planets. Our updated model prefers a larger orbital eccentricity for TOI-199\,c (0.151$\pm$0.003, compared to 0.096$^{+0.008}_{-0.009}$ from \cite{hobson2023}), and a slightly larger mean orbital period for TOI-199\,c, though still within the conservative habitable zone. Our updated TTV model also yields significantly different $\omega$ values for both planets, indicating nearly aligned longitudes of periastron for TOI-199\,b and c, compared to the nearly anti-aligned $\omega$ values from the original fit of \cite{hobson2023}. We see two possible reasons for these changes. The first is simply the addition of new high-precision data improving the accuracy of the TTV fit, especially since there are now 14 total TTV points for a 10-parameter model, which makes the TTV solution less poorly constrained than the original 11 point dataset. The second is the difference in parameterization choice for $e$, $\omega$, and $M_0$ which \cite{hobson2023} samples directly and uses narrow priors defined by initial test runs, but we re-parameterize these quantities using transformations that reduce the correlations between e, $\omega$, and $M_0$ and avoid convergence issues for circular parameters as suggested by previous TTV model comparisons \citep[e.g.,][]{exofast,Agol_2021,Greklek-McKeon2025b}, and we use wide uniform priors with one MCMC sampling until convergence is reached.

The eccentricity of \planetname\ is not large enough to affect its transit shape \citep{hobson2023}, but precise constraints on its $e$ and $\omega$ values may be relevant for future eclipse observations. Given the shift in best-fit $e$ and $\omega$ values when increasing the total data volume from 11 to 14 transits, these best-fit values may still change when more data are available. At the time of this writing, TESS is scheduled to re-observe the TOI-199 system in September, October, and December of 2025, in sectors 96, 97, and 98, respectively. New TTV fits with additional transits from these sectors may be able to further refine the mass of TOI-199\,c and the orbital parameters for both planets. We list the predicted transit times for TOI-199\,b through 2040 December 31 from our best-fit TTV model in Table \ref{tab:transit_times} to aid in planning future observations of this system. Our predicted transit times for planet c do not show any transit events in the TESS data, supporting the conclusion of \citet{hobson2023} that the planet does not transit.

\begin{table}[htbp]
\centering
\caption{Transit Times for TOI-199\,b}
\label{tab:transit_times}
\begin{tabular}{ll}
\hline
\hline
Transit Time (BJD $-$ 2\,457\,000) & Source \\
\hline
$1361.0283^{+0.0008}_{-0.0008}$ & H23 \\
$1465.8841^{+0.0002}_{-0.0002}$ & Predicted \\
$1570.7320^{+0.0010}_{-0.0010}$ & H23 \\
$1675.6182^{+0.0009}_{-0.0009}$ & H23 \\
$1780.5078^{+0.0009}_{-0.0011}$ & Predicted \\
$1885.3868^{+0.0020}_{-0.0026}$ & Predicted \\
$1990.2454^{+0.0019}_{-0.0025}$ & Predicted \\
$2095.1087^{+0.0010}_{-0.0010}$ & H23 \\
$2200.0050^{+0.0006}_{-0.0006}$ & H23 \\
$2304.8772^{+0.0006}_{-0.0006}$ & H23 \\
$2409.7229^{+0.0007}_{-0.0007}$ & H23 \\
$2514.5948^{+0.0051}_{-0.0059}$ & Predicted \\
$2619.4833^{+0.0007}_{-0.0007}$ & H23 \\
$2724.3750^{+0.0040}_{-0.0050}$ & Predicted \\
$2829.2283^{+0.0005}_{-0.0005}$ & H23 \\
$2934.0818^{+0.0003}_{-0.0003}$ & H23 \\
$3038.9735^{+0.0004}_{-0.0004}$ & H23 \\
$3143.8568^{+0.0005}_{-0.0005}$ & TESS Sector 67 \\
$3248.7207^{+0.0083}_{-0.0099}$ & Predicted \\
$3353.5820^{+0.0086}_{-0.0105}$ & Predicted \\
$3458.4552^{+0.0073}_{-0.0091}$ & Predicted \\
$3563.3545^{+0.0064}_{-0.0083}$ & Predicted \\
$3668.2126^{+0.0003}_{-0.0003}$ & JWST \& TESS Sector 87 \\
$3773.0616^{+0.0099}_{-0.0116}$ & Predicted \\
$3877.9412^{+0.0005}_{-0.0005}$ & TESS Sector 94 \\
$3982.8322^{+0.0082}_{-0.0102}$ & Predicted \\
$4087.7172^{+0.0099}_{-0.0125}$ & Predicted \\
$4192.5741^{+0.0094}_{-0.0119}$ & Predicted \\
$4297.4333^{+0.0085}_{-0.0107}$ & Predicted \\
... & ... \\
\hline
\end{tabular}
\tablecomments{H23 is \cite{hobson2023}. This table is truncated for brevity. The full table including predictions of transit times through 2040 December 31, UTC is available in machine-readable form.}
\end{table}

\begin{table}[]
\caption{Prior and posterior distributions from updated TTV modeling including new TESS and JWST data.}
\centering
\makebox[\columnwidth][c]{%
\resizebox{\columnwidth}{!}{%
\begin{tabular}{lcc}
\hline\hline
Parameter & Prior & Posterior \\ \hline
$M_b$ ($M_{\rm J}$) & $\mathcal{N}$(0.169, 0.025) & 0.170 $\pm$ 0.020 \\
$M_c$ ($M_{\rm J}$) & $\mathcal{U}$(0.0, 5.0) & 0.261 $\pm$ 0.005 \\
$P_b$ (days)$^{\star}$ & $\mathcal{U}$(103.8540, 105.8540) & 104.869 $\pm$ 0.001 \\
$P_c$ (days)$^{\star}$ & $\mathcal{U}$(268.690, 278.69) & 274.769$^{+0.144}_{-0.141}$ \\
$\sqrt{e_b}\cos(\omega_b)$ & $\mathcal{U}$(-0.7, 0.7) & 0.136 $\pm$ 0.012 \\
$\sqrt{e_c}\cos(\omega_c)$ & $\mathcal{U}$(-0.7, 0.7) & 0.135 $\pm$ 0.013 \\
$\sqrt{e_b}\sin(\omega_b)$ & $\mathcal{U}$(-0.7, 0.7) & 0.294 $\pm$ 0.008 \\
$\sqrt{e_c}\sin(\omega_c)$ & $\mathcal{U}$(-0.7, 0.7) & 0.364 $\pm$ 0.007 \\
$e_b$ & Derived & 0.105 $\pm$ 0.003 \\
$e_c$ & Derived & 0.151 $\pm$ 0.003 \\
$\omega_b$ & Derived & 65.1 $\pm$ 2.3 \\
$\omega_c$ & Derived & 69.7 $\pm$ 2.1 \\
$T_{0_b}$ (BJD$-$2\,458\,300)$^{\dagger}$ & $\mathcal{U}$(8.5683, 113.3683) & 60.999 $\pm$ 0.001 \\
$T_{0_c}$ (BJD$-$2\,458\,300)$^{\dagger}$ & $\mathcal{U}$(7.43, 281.03) & 160.741$^{+0.685}_{-0.679}$ \\ \hline
\end{tabular}%
}%
}
\label{table:ttv_fit_results}
\begin{tablenotes}
\item[] $\mathcal{N}(a,b)$ is a Gaussian distribution with mean $a$ and standard deviation $b$. The posterior values are reported as the median, with uncertainties given by the 16$^{\textup{th}}$ and 84$^{\textup{th}}$ percentiles.
\item[] $^{\star}$We caution that the period values in this table should never be used to predict future transits in the TOI-199 system. These are approximations of the mean orbital period during the TTV simulation window. Using the period and $T_0$ values listed here to predict transits beyond the simulation window can result in errors much larger than the TTV amplitude shown in Figure~\ref{fig:TTVs}. For transit timing predictions through 2040, see Table~\ref{tab:transit_times}.
\item[] $^{\dagger}$The $T_0$ times listed here are a re-parameterization of the osculating mean anomaly ($M_0$) at the starting time of the simulation (BJD 2\,458\,262.350), and do not represent true transit times.
\end{tablenotes}
\end{table}

\section{Discussion and Conclusions}\label{sec:discussion_and_conclusion}

\subsection{Atmospheric chemistry}

TOI-199\,b occupies a temperature regime that bridges the well-studied populations of hot Jupiters and solar system giants, allowing us to probe distinctive atmospheric chemistry not observed in either extreme. Unlike in hot Jupiters, where thermal chemistry often favors \ce{CO}/\ce{CO2} over \ce{CH4} and \ce{NH3} is largely dissociated, or cold giants, where \ce{CH4} and \ce{NH3} are stable but \ce{H2O} is depleted due to condensation, the chemical composition of TOI-199\,b's atmosphere may provide a more complete census of the basic elemental abundance of the planet. Additional minor species from thermo- and photochemistry reflects a delicate balance between photochemical processes and the strength of vertical mixing.

Taking methane as a proxy for the bulk carbon abundance and thus metallicity \citep{Asplund2009}, our constraints from Table~\ref{tab:exotrpriors} correspond to an atmospheric metallicity of $\ce{C}/\ce{H}=13^{+78}_{-12}\times\,$solar (or $[\ce{C}/\ce{H}]=1.11^{+0.85}_{-1.15}\times\,$solar). However, we consider the high end of this range (i.e., $\gtrsim 50\times$solar) to be disfavored because our photochemical models predict that at high metallicity, \ce{CO} would overtake \ce{CH4} as the dominant carbon carrier, which is inconsistent with our measured upper limits on \ce{CO} and \ce{CO2}. For the same reasoning based on the \ce{CO} upper limit, our models also disfavor internal temperatures much larger than 50 K.

Unlike in cooler gas giants, \ce{H2O} in the atmosphere of TOI-199\,b is expected to be retained in the gaseous phase. Although we lack a direct detection, our \ce{H2O} upper limit constraint is within $1\sigma$ of the \ce{CH4} abundance. Since \ce{H2O} is likely the dominant oxygen carrier in the atmosphere, a measured \ce{CH4}/\ce{H2O} ratio could provide precise constraints on the atmospheric C/O ratio.

The temperate regime of TOI-199\,b makes its nitrogen chemistry especially sensitive to ongoing disequilibrium processes. In thermochemical equilibrium, the dominant nitrogen carrier is predicted to be \ce{NH3}, but with TOI-199\,b's insolation, strong UV irradiation drives active photodissociation of both \ce{NH3} and \ce{N2}, supplying nitrogen for \ce{HCN} formation. This is in contrast to the Jovian atmosphere, where such photochemical processes result in formation of hydrocarbons instead of \ce{HCN} \citep{gladstone1996hydrocarbon, moses2005, hu2021photochemistry}. The strength of vertical mixing determines the relative importance of photochemistry: efficient vertical mixing in the deep atmosphere quenches \ce{NH3} at higher pressures, increasing the supply of \ce{NH3} to the observable atmosphere. Meanwhile, low $K_{\rm zz}$ values in the upper atmosphere, typical of solar system giants, allow photochemistry to dominate, potentially leading to the conversion of most \ce{NH3} to \ce{HCN} in the observable atmosphere. Complicating this interpretation, increased internal temperatures would reduce the abundance of both \ce{NH3} and \ce{HCN}. Thus, a reduced or absent \ce{NH3} abundance could indicate either weak vertical transport or increased internal temperatures. Breaking this degeneracy requires constraints on the \ce{HCN}/\ce{NH3} ratio, together with major oxygen carriers such as \ce{CO} or \ce{CO2}.


In addition, TOI-199\,b's climate allows \ce{OCS} to be more abundant than \ce{SO2}. The basic chemical pathway is similar to that described in \citet{hu2025}. Sulfur originates from the deep atmosphere in the form of \ce{H2S}, which dissociates into \ce{H2} and S. Atomic sulfur then reacts with CO to form \ce{OCS}. Subsequently, the \ce{CH3} radical abstracts sulfur from \ce{OCS} to produce \ce{CH3S}, which acts as a catalyst for the formation of \ce{CH3SH}. Both \ce{OCS} and \ce{CH3SH} remain stable within the observable region of the atmosphere. Although \ce{OCS} is not definitely detected through its 4.8 $\mu$m feature in the current data, our forward model predictions of \ce{OCS} for enhanced metallicity agree well with the retrieved median and upper limit. \ce{OCS} detection in TOI-199\,b may provide a distinctive chemical fingerprint of temperate gas giants that is not observed in the temperature extremes of either solar system giants or hot Jupiters.

\subsection{Follow-up observations}
High-precision observations will be essential to break the degeneracies between metallicity, C/O ratio, and vertical mixing, yielding crucial insights into the formation history and atmospheric dynamics of TOI-199\,b and similar temperate giant planets.

HST GO program 17605 aimed to measure the transmission spectrum of TOI-199\,b with WFC3/G141, targeting \ce{CH4}, \ce{NH3} and \ce{H2O}. However, the HST observation was scheduled based on the period and zero-phase values currently listed on the NASA Exoplanet Archive \citep{hobson2023}, without accounting for TTVs, which caused the observation to miss the transit (private communication). We caution against using the period or $T_0$ values in \citet{hobson2023} or in this paper to schedule future transit observations, since that can result in errors much larger than the TTV amplitude. Instead, we recommend to use Table~\ref{tab:transit_times}, which lists predicted transit times through 2040 December 31.

JWST is currently scheduled to observe two additional transits of TOI-199\,b with NIRSpec to probe the 3--5~$\mu$m region (GO 7188, \citealt{acuna2025}). Section~4 has shown the modeled transmission spectrum of TOI-199\,b for two end-member nitrogen scenarios, anchored by the current data: (i) a Jupiter-like $K_{\rm zz}$ value with 10$\times$ solar metallicity, where \ce{HCN} could be the main nitrogen carrier, and (ii) a hot Jupiter-like $K_{\rm zz}$ value, where \ce{NH3} dominates. \ce{HCN} and \ce{NH3} both absorb near 3.0 $\mu$m, making them indistinguishable in the current data and limiting constraints on $K_{\rm zz}$. However, we perform mock retrievals using our measured NIRSpec/G395H transmission spectrum, scaling the uncertainties down by a factor of six to emulate the precision expected from two transits with no WATA-related systematics. Under these conditions, the abundances of \ce{HCN} and \ce{NH3} could be constrained to $\sim0.2-0.5$ dex precision if they are present in the atmosphere at VMRs of $\sim 10^{-3}$. Such measurements would not only help determine the vertical mixing regime of this planet, but also constrain the photochemical production mechanisms of HCN (from its mixing ratio).

Such data would also constrain species like \ce{CH4} and \ce{CO2} to $\sim0.1$ and $\sim0.2$ dex precision assuming VMRs of $\sim 10^{-2}$ and $\sim 10^{-4}$, respectively. This will enable obtaining a comprehensive picture of the planet's C, N, and O inventories, and therefore precise measurements of the planet's metallicity and the elemental ratios C/O and N/O, which may hold clues to the planet's formation history \citep{oberg2011,turrini2021,pacetti2022}.

If hazes are present in the atmosphere of \planetname, future observations at shorter wavelengths could help distinguish their composition. While the three haze prescriptions explored in this work are all consistent with the current spectrum in $3-5$ $\mu$m, observations spanning $0.6-3\,\mu$m (e.g., with NIRISS/SOSS or NIRSpec's G235 gratings) would help break degeneracies between clouds and tholin/organic hazes through their differing spectral slopes.

The detection of CH$_4$ in TOI-199\,b is consistent with the emerging trend that temperate ($T_{\rm eq} \lesssim 400~K$) low-mean-molecular-weight atmospheres display spectral features in transmission spectroscopy \citep{madhu2023k218b,benneke2024,brande2024,hu2025}, with super-puffs as notable exceptions likely due to high-altitudes hazes and/or circumplanetary rings \citep{alam2022first,libbyroberts2026}. However, TOI-199\,b is the only planet in this list with a mass comparable to that of Jupiter. Ongoing JWST observations of similar long-period planets will not only test whether this trend extends across the gas giant population \citep{fortney2023,cassese2026}, but also begin to constrain their vertical mixing and photochemical processes, paving the way for a new era of atmospheric studies of temperate planets.





%
\section*{Data availability}
The JWST data used in this work are publicly available in the Mikulski Archive for Space Telescopes (MAST) at \dataset[10.17909/wbrk-af94]{https://dx.doi.org/10.17909/wbrk-af94}. The transmission spectra and predicted transit times presented in this work are available on Zenodo: \dataset[10.5281/zenodo.17653116]{https://dx.doi.org/10.5281/zenodo.17653116}

\begin{acknowledgments}
This work is based on observations made with the NASA/ESA/CSA James Webb Space Telescope. The data were obtained from the Mikulski Archive for Space Telescopes at the Space Telescope Science Institute, which is operated by the Association of Universities for Research in Astronomy, Inc., under NASA contract NAS 5-03127 for JWST. These observations are associated with program \#5177. Support for this program was provided by NASA through a grant from the Space Telescope Science Institute under NASA contract NAS 5-03127. Part of this research was carried out at the Jet Propulsion Laboratory, California Institute of Technology, under a contract with the National Aeronautics and Space Administration (80NM0018D0004). Part of the high-performance computing resources used in this investigation were provided by funding from the JPL Information and Technology Solutions Directorate. 
\end{acknowledgments}

\appendix
\setcounter{figure}{0}
\renewcommand{\thefigure}{A\arabic{figure}}
\setcounter{table}{0}
\renewcommand{\thetable}{A\arabic{table}}

\section{Atmospheric Retrievals Supplemental Information}
\begin{figure*}
\centering
\includegraphics[width=1.\linewidth]{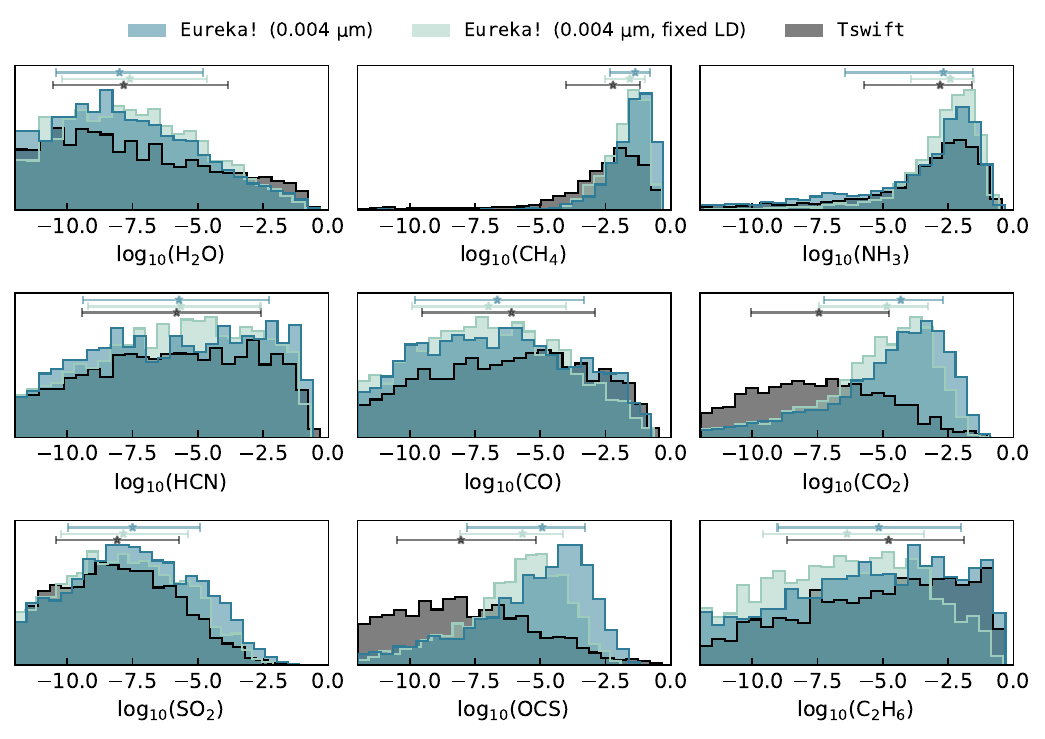}
\caption{Same as Figure~\ref{fig:exotr_posteriors_comparison} but for \texttt{Aurora}.}\label{fig:aurora_posteriors_comparison}
\end{figure*}

Figure~\ref{fig:aurora_posteriors_comparison} shows the VMR posterior distributions from the \texttt{Aurora} retrievals on the different data reductions.

\section{Self-Consistent Model Supplemental Information}\label{app:models}
\begin{figure}
    \centering
    \includegraphics[width=1.\linewidth]{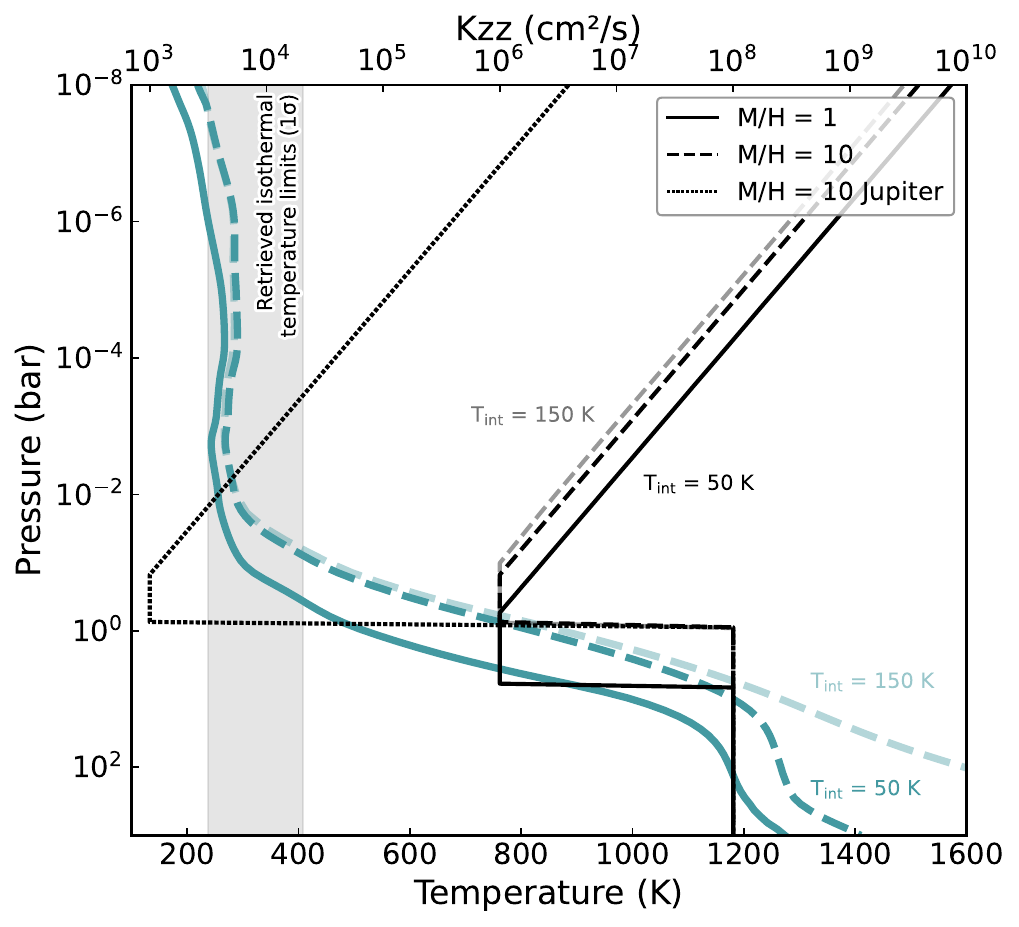}
    \caption{TP profiles (teal curves) and eddy diffusion profiles ($K_{\rm zz}$, black curves) for TOI-199\,b.
    Solid curves represent solar metallicity cases ($\rm{M/H} = 1$), whereas dashed curves are for enhanced metallicity ($\rm{M/H} = 10$). Faint curves represent the case with increased internal temperature. The dotted $K_{\rm zz}$ profile represents a Jupiter-like mixing case. The gray region indicates the retrieved $1\sigma$ temperature limits obtained from \texttt{ExoTR} retrievals.}
    \label{fig:TP_KZZ}
\end{figure}

\texttt{EPACRIS-Climate} solves for the 1D radiative-convective equilibrium using the improved 1D two-stream radiative solver  \citep{Heng_2018_two_stream}. Here, the temperature-pressure (TP) profiles were computed assuming chemical equilibrium abundances for solar and enhanced metallicity cases ($\rm{M/H} = 10$, corresponding to 10 times solar metallicity), as well as for solar and super-solar C/O ratios (0.59 and 1.1). The solar elemental abundances were taken from \citet{Asplund_2021}. We assumed full heat redistribution (f=0.25), a null Bond albedo, and internal temperatures of $T_{\rm int} = 50$~K and $T_{\rm int} = 150$~K. To calculate the lapse rate, we employed the multi-component pseudo-adiabat equation derived by \citet{graham2021multispecies}. We included the opacities of all major expected species: \ce{H2O}, \ce{NH3}, \ce{CH4}, \ce{SO2}, \ce{CO2}, \ce{CO}, \ce{H2S}, \ce{NO2}, \ce{NO}, \ce{N2}, \ce{O2}, \ce{OH}, \ce{HO2}, \ce{HCN}, \ce{OCS}, \ce{O3}, \ce{C2H6}, \ce{CH2O2}, \ce{HNO3}, \ce{N2O}, \ce{C2H2}, \ce{C2H4}, \ce{H2CO}, and \ce{H2O2}, as well as collision-induced absorbers \ce{H2-H2}, \ce{H2-He}, \ce{H2-H}, \ce{H2-N2}, \ce{N2-N2}, and \ce{CO2-CO2}. The sources for the different opacities are the same as those listed in Section~\ref{subsubsec:exotr}. We  used the \texttt{PHOENIX} \citep{Husser2013} stellar spectrum for all forward models. The resulting TP profiles for C/O = 0.59 are shown in Figure~\ref{fig:TP_KZZ}. The TP profiles for C/O = 1.1 did not show substantial differences and are therefore not shown.

Figure~\ref{fig:TP_KZZ} also shows the adopted eddy diffusion coefficient profiles ($K_{\rm zz}$) in our photochemical models. Notionally, we assumed $K_{\rm zz}$ to be constant in the convective region determined by the radiative-convective calculations ($K_{\mathrm{zz}} = 10^{6}~\rm cm^{2}\,s^{-1}$) and the deeper atmosphere (\(K_{\mathrm{zz}} = 10^{8}~ \mathrm{cm}^{2}\,\mathrm{s}^{-1}\)). The convective region value of $10^{6}~ \mathrm{cm}^{2}\,\mathrm{s}^{-1} $ was estimated to be comparable to planets with a similar $T_{\rm eq}$ based on \citet{zhang2018global1}. The deep atmosphere value of $10^8\ \mathrm{cm}^2\,\mathrm{s}^{-1}$ was derived by the mixing-length theory \citep[e.g.,][]{visscher2011quenching} and is consistent with the \ce{GeH4} tracer observations for Jupiter \citep{Wang_2016_Jupiter,bjoraker2018gas}. However, recent work suggests that Jupiter's deep interior $K_{\rm zz}$ should be lower than this value according to 1D-2D coupled approaches \citep{yang2025coupled}, suggesting caution when adopting deep interior $K_{\rm zz}$ values. For comparison, we also ran models with a Jupiter-like $K_{\rm zz}$ profile, where $K_{\mathrm{zz}}$ is reduced to $10^{3}~\rm\,cm^{2}\,s^{-1}$ in the convective region \citep{gladstone1996hydrocarbon}. For both cases, $K_{\rm zz}$ was assumed to increase with the inverse square root of pressure above the convective region \citep{lindzen1981turbulence} as implemented in \citet{hu2021photochemistry}. Finally, for further comparison, we consider another set of cases where a uniform $K_{\rm zz}$ is adopted throughout the atmosphere, with values ranging from $10^4$ to $10^8\ \mathrm{cm}^2\,\mathrm{s}^{-1}$, as these are typical assumptions for hot Jupiter studies \citep{zhang2018global1}.


Using the TP and $K_{\rm zz}$ profiles described above, we ran a suite of photochemical simulations with \texttt{EPACRIS-Chemistry} \citep{Yang2024_epacris}, adopting the chemical network of \citet{Yang2024_subneptunes, hu2025, yang_2025_16750016} (99 species and 2024 reactions in total), and the same \texttt{PHOENIX} stellar spectrum as described previously. We note that \texttt{PHOENIX} models typically underestimate UV flux \citep{Behr_2023}, which may lead to slightly conservative estimates of photodissociation rates affecting photochemically produced species.

To generate the corresponding transmission spectra, we used the radiative-transfer code \texttt{petitRADTRANS}\footnote{\url{http://gitlab.com/mauricemolli/petitRADTRANS}; version 2 was used for this work} \citep{Molliere_2019,Molliere_2020}. Spectra were computed at a resolving power of $\lambda/\Delta\lambda = 1000$; although the figures display binned versions for clarity, the full‐resolution spectra were used to compute $\chi^{2}$ values for all forward models. For the reference pressure defining the radius of the planet we used $P_{0} = 1$ bar. Most molecular opacities in \texttt{petitRADTRANS} were taken from the ExoMol database\footnote{\url{https://www.exomol.com/}} \citep{Chubb_2021}. A complete list of opacities and their line-list sources is provided in Table A.1 of \citet{Zilinskas_2025}.

\bibliography{sample701}{}
\bibliographystyle{aasjournalv7}



\end{document}